\documentclass[a4paper,11pt]{article}
\usepackage{jcappub} % for details on the use of the package, please see the JINST-author-manual
\usepackage{lineno}
\usepackage{tikz}
\usepackage{slashed}

\usetikzlibrary{decorations.markings}
\usetikzlibrary{decorations.pathmorphing}
\tikzstyle{fermion} = [very thick, style={decoration={markings,mark=at position .5 with {\arrow{>}}},postaction={decorate}}]
\tikzstyle{gauge} = [very thick, style={decorate, decoration=snake}]
\tikzstyle{scalar} = [very thick, dashed]

\title{Relaxing Limits from Big Bang Nucleosynthesis on Heavy Neutral Leptons with Axion-like Particles}

\newcommand{\AddrUCL}{Department of Physics and Astronomy, University College London,\\London WC1E 6BT, United Kingdom}

\newcommand{\AddrKIT}{Institute for Theoretical Particle Physics (TTP), Karlsruhe Institute of Technology (KIT), 76128 Karlsruhe, Germany}

\affiliation[a]{\AddrUCL}
\affiliation[b]{\AddrKIT}

\author[a]{Frank F. Deppisch,} 
\emailAdd{f.deppisch@ucl.ac.uk}
\author[b]{Tom\'as E. Gonzalo,} 
\emailAdd{tomas.gonzalo@kit.edu}
\author[a]{Chayan Majumdar,} 
\emailAdd{c.majumdar@ucl.ac.uk}
\author[a]{Zhong Zhang} 
\emailAdd{zhong.zhang.19@ucl.ac.uk}

\abstract{Heavy neutral leptons (HNLs) are constrained by requirements of Big Bang Nucleosynthesis (BBN) as their decays significantly impact the formation of the primordial elements. We propose here a model where the primary decay channel for the HNLs is to an axion-like particle (ALP) and a neutrino. Consequently, HNLs can decay earlier and evade the BBN bound for lower masses, provided the ALPs themselves decay considerably later. Further cosmological and astrophysical constraints limit severely the range of validity of the ALP properties. We find that a new parameter region opens up for HNLs with masses between 1~MeV and 1~GeV, and active-sterile neutrino mixing strengths between $10^{-9}$ and $10^{-6}$ that is consistent with constraints and can be probed in future searches. In such a scenario, current bounds as well as sensitivities of future direct HNL searches such as at NA62 and DUNE will be affected.}

\begin{document}
\maketitle
\flushbottom

\section{Introduction}

Heavy Neutral Leptons (HNLs) are amongst the most popular exotic particles thought to exist beyond the particle content of the Standard Model (SM). Assumed to be fermions that are neutral under the SM gauge symmetries, they act as heavy right-handed neutrinos and, if they are of Majorana nature, trigger a seesaw mechanism of light neutrino mass generation \cite{Minkowski:1977sc,Mohapatra:1979ia,Gell-Mann:1979vob, Yanagida:1979as, Schechter:1980gr}. In this context, they may also help to understand the matter-antimatter asymmetry of the universe, as their $CP$-violating decays or oscillations can generate a baryon asymmetry through leptogenesis \cite{Fukugita:1986hr}. HNLs may thus help to explain why the active neutrinos are so much lighter than the other SM fermions, but not massless as required from observing oscillations \cite{ParticleDataGroup:2024cfk, Esteban:2020cvm}, and the origin of matter in the universe.

In the minimal scenario, an HNL $N$ couples to the SM only through its mixing $U_{\alpha N}$ ($\alpha = e,\mu,\tau$) with the active neutrinos, induced by the Yukawa coupling of the HNL with the SM lepton doublets $L_\alpha$ and Higgs doublet $H$. The HNL then also participates in charged and neutral lepton currents. There are numerous past, current and planned future searches for HNLs based on this, over a wide range of HNL masses $m_N$; from the eV-scale where HNLs can be tested in oscillations, over keV and MeV scales mainly probed in nuclear processes such as $\beta$ decay, MeV to GeV in beam dump and meson decays, to electroweak scale masses and above probed in colliders. A recent overview of current and future searches is provided in~\cite{Bolton:2019pcu,Abdullahi:2022jlv} and a global study of current constraints on HNLs can be found in \cite{Chrzaszcz:2019inj}. The mass range $100~\text{MeV} \lesssim m_N \lesssim 100$~GeV has especially been targeted in searches for HNLs, since in this mass range HNLs are naturally long-lived for the small active-sterile mixing strengths expected for successful neutrino mass generation. This results in macroscopic decay lengths and thus displaced decay vertices that can be looked for with a high sensitivity. 

Being long-lived, HNLs also affect the early history of the universe beyond their role in leptogenesis. If HNLs had been in thermal equilibrium, and they decay around or later than about a second after the big bang, the produced particles affect Big Bang Nucleosynthesis (BBN) \cite{Boyarsky:2021yoh, GAMBITCosmologyWorkgroup:2020htv}. In the absence of other channels, the primary decay of HNLs to SM particles is hadronically \cite{Atre:2009rg, Coloma:2020lgy}. The mesons produced through these decays interact strongly with the protons and neutrons of the primordial plasma, thereby modifying the $p\leftrightarrow n$ conversion ratio which sets the primordial abundances. Most notably, the primordial abundance of $^4$He, precisely measured as $Y_p = 0.2436 \pm 0.0038$~\cite{Hsyu:2020uqb}, would be mostly affected since most primordial neutrons and protons end up forming $^4$He. Such considerations disfavour HNLs with masses $m_N \lesssim 1$~GeV, for small active-sterile mixing strengths required by current constraints and expected for light neutrino mass generation. Even lighter and longer-lived $N$ are likewise disfavoured as they will act as additional degrees of freedom, or inject them through their decays, and may overclose the universe.

The above considerations apply for the minimal scenario where the HNL only couples via the active-sterile mixing. Instead, HNLs may also interact with an expanded exotic sector within specific beyond-the-SM scenarios. Often, these are high-scale scenarios where the HNL is charged under an additional gauge force, such as $B-L$ \cite{Davidson:1978pm, Marshak:1979fm, Mohapatra:1980qe, Davidson:1987mh, Buchmuller:1991ce} and left-right-symmetric \cite{Pati:1973rp, Pati:1973uk, Pati:1974yy, Mohapatra:1974gc, Senjanovic:1975rk, Mohapatra:1980qe} models. Less explored is the possibility of coupling the HNL to a lighter exotic sector. We here consider such a scenario where the HNL couples to a light, exotic pseudoscalar $a$, akin to an axion-like particle (ALP) \cite{Marsh:2015xka, Chadha-Day:2021szb, Marsh:2023tep}. Recent works exploring similar HNL-ALP portals in colliders can be found in \cite{deGiorgi:2022oks, Marcos:2024yfm,Wang:2024mrc, Wang:2024prt} and \cite{Arguelles:2021dqn, Wang:2024mrc} in beam dump experiments. We will refer to our dark scalar as ALP in the following, and it is dark in the sense that it couples only to the HNL in the first instance, with interactions to the SM suppressed by the active-sterile mixing and through loops. 

Our motivation is to explore the cosmological consequences of changing the HNL decay width due to the additional channel $N\to a\nu$ in the scenario. While suppressed by both the active-sterile neutrino mixing and the ALP decay constant $f_a$, it is a two-body decay that can compete with the three-body (at parton level) decays to SM particles only. This will enlarge the region of interest constrained by BBN, i.e., where HNLs decay earlier and motivate direct searches under such an additional invisible HNL decay. At the same time, the production of ALPs and their own decays will themselves lead to constraints on the viable parameter space by ensuring the cosmological history, especially until the time of the formation of the cosmic microwave background (CMB), is not affected. Any overabundance of ALPs would modify the expansion history of the Universe, either by also inducing a change on the primordial element abundances, if they are efficiently produced before BBN, or by increasing the temperature of the neutrino bath, if they decay strongly to active neutrinos before recombination. Furthermore, ALPs may be produced in the core of stars or supernovae, which leads to faster cooling and anomalous neutrino fluxes, posing additional constraints on strongly coupled ALPs.

The paper is organized as follows. In Sec.~\ref{sec:model}, we describe the model and determine the HNL and ALP decay widths as important quantities for our later considerations. In Sec.~\ref{sec:cosmo}, we describe our modelling of the cosmological history in our scenario by setting up the Boltzmann equations which we separate into a regime before and after BBN. The relevant constraints from cosmology and astrophysics, as well as the expectations for direct searches, are then discussed in Sec.~\ref{sec:constraints}, and they are applied in Sec.~\ref{sec:results} where we present our results in terms of the viable parameter space. Lastly we provide a conclusion and outlook in Sec.~\ref{sec:conclusions}.

\section{Model}
\label{sec:model}

In this work, we have extended the existing SM particle spectrum by adding a SM gauge singlet HNL and an ALP. The ALP only couples to the HNL in the unbroken SM and thus the mass of the ALP is not strictly tied to the QCD phase transition scale, $\Lambda_\text{QCD}$. In this extension, we outline a phenomenological framework which can describe the coupling between HNL and ALP and its implications for current phenomenological, cosmological and astrophysical constraints from other complimentary studies.     

\subsection{Lagrangian}

In order to generate at least two non-degenerate active neutrino masses, as confirmed by the results from various neutrino oscillation experiments, it is necessary to introduce at least two HNLs. In general, we can extend the SM to include $\mathcal{N}$ HNLs as SM gauge-singlet Weyl fermion fields $N_{iR}$ ($i = 1, .., \mathcal{N}$). Including a pseudoscalar ALP $a$ that couples solely to $N_{iR}$, the most general and renormalisable Lagrangian is
\begin{align}
    \mathcal{L} 
    = \mathcal{L}_\text{SM} 
    + i\bar{N}_{iR}\slashed{\partial}N_{iR} 
    - (Y_\nu)_{\alpha i}\bar{L}_\alpha \tilde{H} N_{iR} 
    - \frac{1}{2}(\mathcal{M}_R)_{ij}\bar{N}_{iR}^c N_{jR} 
    + \mathcal{L}_{aNN} + \text{h.c.}
\end{align}
Here, $L_\alpha = (\nu_{\alpha L}, \ell_{\alpha L})^T$ and $H = (H^0, H^-)^T$ are the SM lepton and Higgs doublets, with flavour index $\alpha = e, \mu, \tau$ and $\tilde{H} \equiv i \sigma_2 H$ as the dual of $H$. $Y_\nu$ corresponds to the general complex Yukawa coupling matrix which couples $N_{iR}$ to the SM spectrum. As HNLs are electromagnetically neutral and SM gauge-singlet particles, one can write a Majorana mass term where $\mathcal{M}_R$ corresponds to the lepton-number breaking scale. After electroweak symmetry breaking, the Higgs doublet acquires a vacuum expectation value $v/\sqrt{2} = \langle H \rangle$ which correspondingly generates the Dirac mass for neutrinos in the theory as $(\mathcal{M}_D)_{\alpha i} = (Y_\nu)_{\alpha i} \frac{v}{\sqrt{2}}$. The neutral fermion mass matrix can be written in the $(\nu_{\alpha L}^c, N_{jR})^T$ basis by a $(3+\mathcal{N}) \times (3+\mathcal{N})$ complex symmetric matrix as
\begin{align}
    \mathcal{M}_\nu = 
    \begin{pmatrix}
      0               & \mathcal{M}_D \\
      \mathcal{M}_D^T & \mathcal{M}_R
    \end{pmatrix},
\end{align}
where $\mathcal{M}_D$ and $\mathcal{M}_R$ are 3$\times \mathcal{N}$ and $\mathcal{N} \times \mathcal{N}$ matrices, respectively. We can consider $\mathcal{M}_R$ as diagonal without any loss of generality. One can introduce a $(3+\mathcal{N}) \times (3+\mathcal{N})$ unitary matrix,
\begin{align}
    U = \begin{pmatrix}
        U_{\nu\nu} & U_{\nu N} \\
        U_{N \nu} & U_{NN}
    \end{pmatrix},
\end{align}
which diagonalizes the matrix $\mathcal{M}_\nu$. $U_{\nu\nu}$ is related to the $3\times3$ Pontecorvo-Maki-Nakagawa-Sakata (PMNS) matrix, but may include non-unitary corrections due to the presence of $3\times \mathcal{N}$ active-sterile mixing $U_{\nu N} = U_{N \nu} \equiv \mathcal{M}_D \mathcal{M}_R^{-1}$. Using this unitary matrix, the diagonalized neutral fermion mass matrix in the Majorana mass eigenbasis $(\nu_\lambda, N_\kappa)^T (\lambda = 1, 2, 3; \kappa = 1, .., \mathcal{N})$ will be
\begin{align}
    U^\dagger \mathcal{M}_\nu U^\ast = U^\dagger \cdot 
    \begin{pmatrix}
        0               & \mathcal{M}_D \\
        \mathcal{M}_D^T & \mathcal{M}_R
    \end{pmatrix} \cdot U^\ast 
    = \begin{pmatrix}
        (m_\nu)_{3\times 3} & 0  \\
        0                   & (M_{N})_{\mathcal{N} \times \mathcal{N}}
    \end{pmatrix}.
\end{align}
Here, $m_\nu$ and $M_{N}$ are the diagonalized active and heavy neutrino masses, respectively. In the mass basis, the Majorana mass matrices can be written as
\begin{align}
    m_\nu & \approx -\mathcal{M}_D \mathcal{M}_R^{-1} \mathcal{M}_D^T, \nonumber\\
    M_N   & \approx \mathcal{M}_R,
\end{align}
using the type-I seesaw approximation $\mathcal{M}_D \ll \mathcal{M}_R$.

In addition to the regular seesaw sector, we also assume that the ALPs $a$ couples with the HNL $N$, but not with other SM particles directly. Hence, due to the pseudoscalar nature of ALPs, the Lagrangian for the ALP-HNL interaction is given by the derivative coupling~\cite{Gola:2021abm} in the mass basis as
\begin{align}
    \mathcal{L}_{aNN} 
    = \sum_{\kappa = 1}^\mathcal{N} \frac{1}{f_a}(\partial_\mu a) 
      \Bar{N}_\kappa \gamma^\mu \gamma_5 N_\kappa 
    = -\sum_{\kappa = 1}^\mathcal{N} \frac{2i}{f_a} m_{N_\kappa} a 
      \Bar{N}_\kappa \gamma_5 N_\kappa, 
\label{eq:LNN}
\end{align}
where $f_a$ is the ALP decay constant and $m_{N_\kappa}$ are the eigenvalues of $M_N$. The second equality holds by applying the equation of motion for the HNLs and removing a total derivative after integration by parts~\cite{Gelmini:1982zz}. Due to the small active-sterile mixing $U_{\nu N} = U_{N\nu} \equiv \sqrt{m_\nu M_N^{-1}}$, an interaction between the ALP and the active neutrinos $\nu_\lambda$ is induced. 
In the mass basis of the active neutrinos and HNLs, we can write the interaction Lagrangian between the ALP and active neutrinos as
\begin{align}
    \mathcal{L}_{a\nu\nu} & = -\sum_{\lambda, \lambda^\prime=1}^3 \sum_{\kappa =1}^{\mathcal{N}}\frac{2i}{f_a} m_{N_{\kappa}} a (U_{\nu \nu}^\dagger U_{N \nu}^T)_{\kappa\lambda} (U_{N \nu}^\ast U_{\nu\nu})_{\kappa\lambda^\prime} \bar{\nu}_{\lambda} \gamma_5 \nu_{\lambda^\prime}\nonumber \\
    & \approx -\sum_{\lambda, \lambda^\prime=1}^3 \frac{2i}{f_a} m_{N} a |U_{N \nu}U_{\nu\nu}^\ast|^2_{\lambda\lambda^\prime}  \bar{\nu}_{\lambda} \gamma_5 \nu_{\lambda^\prime}
    \label{eq:Lnunu}
\end{align}
where the second expression applies for (nearly) degenerate HNLs and $\lambda, \lambda^\prime$ denote the active neutrino mass eigenstates. Furthermore, the interaction Lagrangian for the $aN\nu$ vertex can be written similarly as
\begin{align}
    \mathcal{L}_{aN\nu} = 
    -\sum_{\lambda^\prime = 1}^3 \sum_{\kappa = 1}^{\mathcal{N}} 
    \frac{2i}{f_a}m_{N_\kappa} a (U_{N\nu}^\ast U_{\nu\nu})_{ \kappa\lambda^\prime}\bar{N}_\kappa \gamma_5 \nu_{\lambda^\prime}.
\end{align}

While at least two HNLs are needed to explain all active neutrino masses and mixing, we are mainly interested in elucidating the principle effects in our framework. For simplicity, by considering a single HNL, $N_1 \equiv N$, we can acquire all the important information from the framework using this simplified viewpoint. Furthermore, we consider the active-sterile mixing of $N_1$ with a single active neutrino $\nu_1$ which is mostly electron-type neutrino i.e., $\nu_1 \equiv \nu_e$, without loss of generality. From now on, we take $|U_{\nu\nu}|^2 \sim \mathcal{O}(1) \gg |U_{\nu N}|^2 = |U_{N\nu}|^2$ and $U_{\nu N} = U_{eN}$. Equations~\eqref{eq:LNN} and \eqref{eq:Lnunu} then take the form
\begin{align}
    \mathcal{L}_{aNN}     & = -\frac{2i}{f_a} m_N a\bar{N}\gamma_5 N, \nonumber\\
    \mathcal{L}_{aN\nu}   & = -\frac{2i}{f_a} m_N U_{eN} a\bar{N}\gamma_5 \nu_e,
    \nonumber\\
    \mathcal{L}_{a\nu\nu} & = -\frac{2i}{f_a} m_N |U_{eN}|^2 a\bar\nu_e\gamma_5\nu_e 
                            = -\frac{2i}{f_a} m_\nu a\bar\nu_e \gamma_5 \nu_e,
\label{eq:Lsimplified}
\end{align}
with the light neutrino mass $m_\nu = |U_{eN}|^2 m_N$ induced by the seesaw mechanism. 

\subsection{HNL decays}
\label{subsec:HNL_decays}

A Majorana HNL with a mass of a few hundred MeV to GeV decays via various channels into SM particles. All these decays are four-fermion interactions mediated by either a $Z$ or $W$ boson. We will briefly discuss these SM decay channels of HNLs here with a detailed discussion found in \cite{Atre:2009rg, Coloma:2020lgy}. In our framework with only one HNL species that couples with the first generation only, we have the corresponding decay channels of HNLs as:

\paragraph{$N\to \nu_e \ell^-\ell^+$:} mediated by charged (for $\ell = e$) and neutral currents (for $\ell = e,\mu,\tau$). The corresponding decay width is (with $x_\ell = m_\ell/m_N$)
\begin{align}
    \Gamma^{\nu_e\ell^-\ell^+} 
    = |U_{e N}|^2 \frac{G_F^2 m_N^5}{96\pi^3}
      \left[\left(C_1 + 2\sin^2\theta_W \delta_{e,\ell}\right) f_1(x_\ell)
    + \left(C_2 + \sin^2\theta_W \delta_{e,\ell}\right) f_2 (x_\ell)\right],
\end{align}
with
\begin{align}
    C_1 = \frac{1}{4}(1-4\sin^2\theta_W + 8\sin^4\theta_W), \qquad 
    C_2 = \frac{1}{2}(-\sin^2\theta_W + 2 \sin^4 \theta_W),
\end{align}
and the functions are defined as
\begin{align}
    f_1(x) &= (1-14x^2-2x^4-12x^6)\sqrt{1-4x^2} + 12x^4(x^4-1)L(x) \nonumber\\
    f_2(x) &= 4\left[x^2(2+10x^2-12x^4)\sqrt{1-4x^2} + 6x^4(1-2x^2+2x^4)L(x)\right],
\end{align}
with 
\begin{align}
    L(x) = \ln\left[\frac{1-3x^2-(1-x^2)\sqrt{1-4x^2}}{x^2(1+\sqrt{1-4x^2})}\right].
\end{align}

\paragraph{$N\to e^-\ell^+\nu_\ell$:} mediated by a charged current with $\ell = \mu, \tau$. The decay width for this process is
\begin{align}
    \Gamma^{e^-\ell^+\nu_\ell} 
    = |U_{e N}|^2 \frac{G_F^2 m_N^5}{192\pi^3}
    \left[1-8x_\ell^2 + 8x_\ell^6 - x_\ell^8 -12x_\ell^4 \ln(x_\ell^2)\right].
\end{align}

\paragraph{$N \to \nu_e \nu_\ell \bar{\nu}_\ell$:} mediated by a neutral current with $\ell = e,\mu,\tau$. The decay width can be written as
\begin{equation}
    \Gamma^{\nu_e\nu_\ell\bar{\nu}_\ell} 
    = \frac{G_F^2}{96\pi^3}|U_{eN}|^2 m_N^5.
\end{equation}

\paragraph{$N \to P \nu_e$:} with a neutral pseudoscalar meson $P = \pi^0, K^0, \eta, \eta^\prime$. The decay width is
\begin{align}
    \Gamma^{P \nu_e} = \frac{G_F^2 m_N^3}{32\pi}f_P^2 |U_{eN}|^2 (1-x_P^2)^2,
\end{align}
with the meson decay constant $f_P$ whose values we have taken from \cite{Coloma:2020lgy} and $x_P = m_P / m_N$.

\paragraph{$N \to P^+ e^-$:} with a charged pseudoscalar meson $P^+ = \pi^+, K^+, D^+, D_s^+$. The decay width is
\begin{align}
    \Gamma^{P^+ e^-} 
    = \frac{G_F^2 m_N^3}{16\pi}f_P^2 |U_{eN}|^2 |V_{qq^\prime}|^2 
    \lambda^{1/2}(1,x_P^2,x_e^2)\left[1-x_P^2-x_e^2(2+x_P^2-x_e^2)\right].
\end{align}
where $V_{qq^\prime}$ denotes the CKM mixing matrix element involving relevant quarks in the produced meson and $\lambda (a,b,c) \equiv (a-b-c)^2-4bc$.

\paragraph{$N \to V \nu_e$:} with a neutral vector meson $V = \rho, \omega, \phi, K^{*0}$. The decay width can be written as
\begin{align}
    \Gamma^{V \nu_e} 
    = \frac{G_F^2 m_N^3}{32\pi m_V^2} f_V^2 \kappa_V^2 |U_{eN}|^2 (1+2x_V^2)(1-x_V^2)^2.
\end{align}
where $f_V$ and $\kappa_V$ represent the decay constant and the vector coupling associated with the produced neutral vector mesons, respectively.

\paragraph{$N \to V^+ e^-$:} with a charged vector meson $V^+ = \rho^+, K^{\ast,+}$. The decay width can be written as
\begin{align}
    \Gamma^{V^+ e^-} 
    = \frac{G_F^2 m_N^3}{16\pi m_{V^\pm}^2} f_V^2 |U_{eN}|^2 |V_{qq^\prime}|^2 \lambda^{1/2}(1,x_V^2, x_e^2)\left[(1-x_V^2)(1+2x_V^2)+x_e^2(x_V^2+x_e^2-2)\right].
\end{align}

The total decay width for a Majorana HNL decaying into purely SM particles is then
\begin{align}
    \Gamma^{N \to \text{SM}} &= 
       \sum_{\ell} \Gamma^{\nu_e \ell^-\ell^+} 
     + \sum_{\ell = \mu, \tau} 2\Gamma^{e^- \ell^+ \nu_\ell} 
     + \sum_{\ell} \Gamma^{\nu_e \nu_\ell \bar{\nu}_\ell} \nonumber\\
    &+ \sum_P \Gamma^{P \nu_e} 
     + \sum_P 2\Gamma^{P^+ e^-}  
     + \sum_V \Gamma^{V \nu_e} 
     + \sum_V 2\Gamma^{V^+ e^-}.
\end{align}
The factors of two are due to the Majorana HNL decaying into opposite charge combinations, e.g., $\Gamma^{P^- e^+} = \Gamma^{P^+ e^-}$.

Due to the interaction in Eq.~\eqref{eq:Lsimplified}, a new decay channel for HNLs is also possible, where the HNL decays as $N \to a\nu$ via the active-sterile mixing at tree level. The decay width, in the limit where $m_\nu \ll m_N$, can be written as
\begin{align}
    \Gamma^{N \to a\nu} 
    = \frac{|U_{eN}|^2 m_N^3}{4\pi f_a^2}
      \sqrt{1+\left(\frac{m_a}{m_N}\right)^2} 
      \left[1-\left(\frac{m_a}{m_N}\right)^2\right]^{3/2} 
    \approx \frac{|U_{e N}|^2 m_N^3}{4\pi f_a^2},
\label{eq:Nanu}
\end{align}
where the last equality holds in the limit where $m_a, m_\nu \ll m_N$. Hence the lifetime of the HNL considering only this decay mode, calculated in its rest frame, can be expressed as
\begin{align}
    \tau_{N \to a\nu} \approx 8.6 \times 10^{-4}~ \text{s} 
    \times \left(\frac{f_a}{1~\text{TeV}}\right)^2 
    \times \left(\frac{10^{-14}}{|U_{e N}|^2}\right) 
    \times \left(\frac{1~\text{GeV}}{m_N}\right)^3.
\end{align}
The total HNL decay width to SM particles can be approximately written as \cite{Bolton:2019pcu}
\begin{align}
    \Gamma^{N\to \rm{SM}} &\approx \left(30~ \Gamma^{2-\rm{body}} + 10~ \Gamma^{3-\rm{body}}\right) |U_{eN}|^2 \nonumber\\
    &\approx \left(\frac{6 f_M^2}{\pi} + \frac{m_N^2}{20\pi^3}\right) m_N^3 |U_{eN}|^2 G_F^2,
\end{align}
where $f_M \approx \mathcal{O}(0.1)$~GeV corresponds to the typical decay constant of the produced pseudoscalar or vector meson. Comparing this with Eq.~\eqref{eq:Nanu}, it is clear that the $N\to a\nu$ channel will be dominant if
\begin{equation}
    m_N < \frac{\sqrt{5|1-24G_F^2f_M^2f_a^2|}\pi}{f_a G_F},
    \label{eq:ALPdecaydomination}
\end{equation}
independent of $|U_{eN}|^2$.

\begin{figure}[t!]
    \centering
    \includegraphics[width=0.49\textwidth]{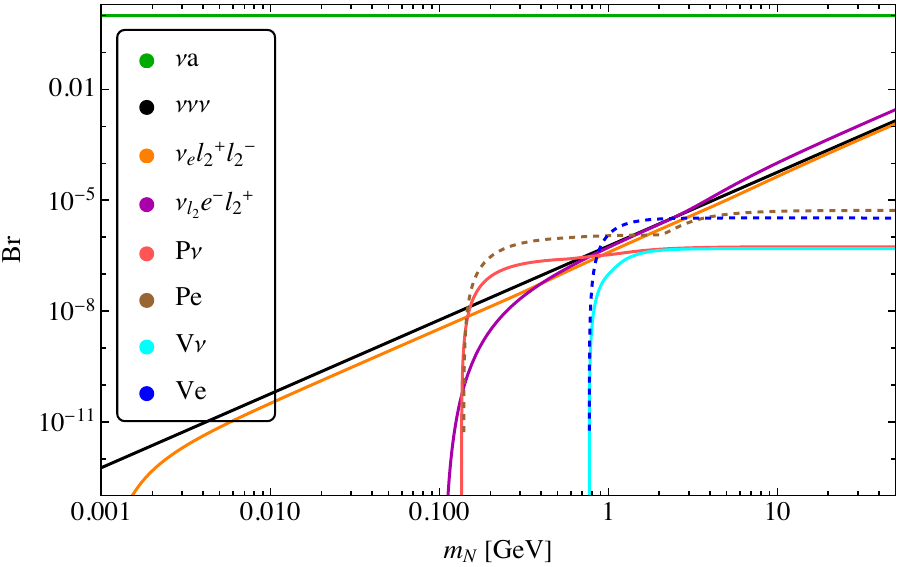}
    \includegraphics[width=0.49\textwidth]{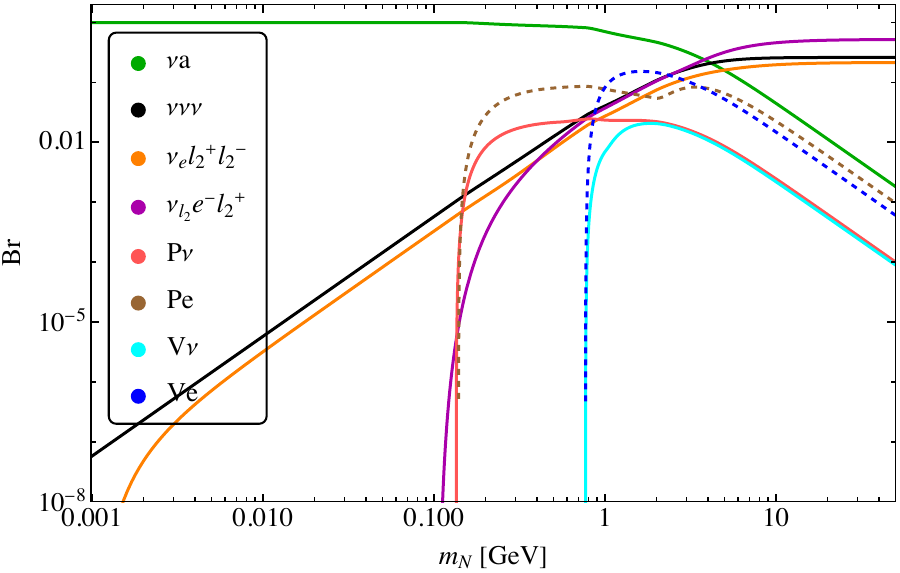}
    \caption{Branching ratios of HNL as a function of the HNL mass $m_N$ for an ALP mass $m_a = 1$~keV, and decay constant $f_a = 1$~TeV (left panel) and $f_a = 10^{2.5}$~TeV (right panel).}
    \label{fig:brN}
\end{figure}
The branching ratios of the HNL decaying to various particles as a function of mass of the HNL for two different values of $f_a$ are shown in Fig.~\ref{fig:brN}. In this work, as we are mostly interested in the regime where $m_a \ll m_N$, correspondingly we have chosen $m_a = 1$ keV and varied $m_N \in [10^{-3}-50]$ GeV range. In the left panel, we have considered $f_a = 1$ TeV where the Br$(N \rightarrow a \nu) \approx 100 \%$ (solid green) throughout the entire mass range of HNL, while the behaviour of different SM decay channels are shown as, $N \rightarrow \nu\nu\nu$ (solid black), $N \rightarrow \nu_e (e^+ e^- + \mu^+ \mu^- + \tau^+ \tau^-)$ (solid orange), $N \rightarrow e^- (\nu_\mu \mu^+ + \nu_\tau \tau^+)$ (solid magenta), different pseudoscalar ($P$) and vector ($V$) mesonic channels (which become relevant for $m_N \geq m_\pi$) as $N \rightarrow P \nu$ (solid red), $N \rightarrow P e$ (dashed brown), $N \rightarrow V \nu$ (solid cyan), $N \rightarrow V e$ (dashed blue). In the right panel of the figure we have taken $f_a = 10^{2.5}$ TeV, for which the coupling of the ALP to HNL significantly decreases. Here, for lower mass of HNL the axionic decay channel still dominates the scenario, as expected from eq.~\eqref{eq:ALPdecaydomination}, while for $m_N > m_\pi$, as the mesonic channels come into picture, the branching ratio to axionic channel drops significantly as compared to the SM decay channels.           

\subsection{ALP decays}

The interaction between the ALPs and active neutrinos in eq. \eqref{eq:Lnunu} causes the ALP to decay. This decay channel occurs at tree level, thus dominating the ALP decay width, and it is given by
\begin{equation}
    \label{eq:anunu}
    \Gamma^{a \rightarrow \nu\nu} = \frac{1}{f_a^2} \frac{m_N^2 m_a U_{e N}^4}{2\pi} \sqrt{1-\frac{4m_\nu^2}{m_a^2}}\left(1-\frac{2m_\nu^2}{m_a^2}\right) \simeq \frac{ m_N^2 m_a U_{e N}^4}{2\pi f_a^2}
\end{equation}
where we have considered $m_a \gg m_\nu$ to arrive at the last expression. 

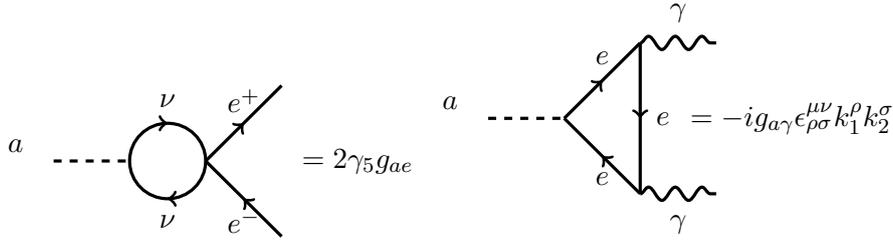
\begin{figure}[t!]
 \centering
 \begin{tikzpicture}[text centered]
  \draw [scalar]  (-1,0) -> (0,0);
  \node at (-1.5,0.2) {$a$};
  \draw [fermion]  (0,0) arc (180:0:0.5););
  \node at (0.5,0.8) {$\nu$};
  \draw [fermion]  (1,0) arc (0:-180:0.5););
  \node at (0.5,-0.8) {$\nu$};
  \draw [fermion] (1,0) -> (2,1);
  \node at (1.5, 0.8) {$e^+$};
  \draw [fermion] (2,-1) -> (1,0);
  \node at (1.5,-0.8) {$e^-$};
  \node at (3,0) {$ = 2 \gamma_5 g_{ae}$};
 \end{tikzpicture}
 \begin{tikzpicture}[text centered]
  \draw [scalar]  (-1,0) -> (0,0);
  \node at (-1.5,0.2) {$a$};
  \draw [fermion]  (0,0) -> (1,1);
  \node at (0.5,0.8) {$e$};
  \draw [fermion]  (1,-1) -> (0,0);
  \node at (0.5,-0.8) {$e$};
  \draw [fermion] (1,1) -> (1,-1);
  \node at (1.3, 0) {$e$};
  \draw [gauge] (1,1) -> (2,1);
  \node at (1.5,1.4) {$\gamma$};
  \draw [gauge] (1,-1) -> (2,-1);
  \node at (1.5,-1.4) {$\gamma$};
  \node at (3,0) {$ = -ig_{a\gamma}\epsilon^{\mu\nu}_{\rho\sigma}k_1^\rho k_2^\sigma$};
 \end{tikzpicture}
 \caption{Effective couplings of the ALP to electrons (left) and photons (right).}
 \label{fig:effective_couplings}
\end{figure}

No other decay channels are available at tree level. However, as it will be seen below (see section \ref{sec:constraints}), most of the constraints on axion-like particles come from their interactions with electrons and photons. Though the Lagrangian in eq.~\eqref{eq:Lsimplified} does not produce axion-electron or axion-photon interactions at tree level, such couplings can be induced at 1-loop and 2-loop respectively, as shown in Figure \ref{fig:effective_couplings}. 

These effective axion-electron and axion-photon couplings, $g_{ae}$ and $g_{a\gamma}$, can cause the ALP to decay to a pair of electrons or photons and, in the limit where $m_a, m_\nu, m_e \ll m_N$, are given by
\begin{align}
    g_{ae} &\approx \frac{\sqrt{2}G_F g_{aN}|U_{e N}|^4 m_em_N}{16\pi^2} = \frac{\sqrt{2}G_F|U_{e N}|^4 m_em_N^2}{16\pi^2f_a}, \notag \\
    g_{a\gamma} &\approx \frac{e^2 g_{ae}}{2\pi^2 m_e}\left(1+\frac{1}{12}\frac{m_a^2}{m_e^2}\right) = \frac{ \sqrt{2}e^2 G_F|U_{e N}|^4 m_N^2}{32\pi^4 f_a}\left(1+\frac{1}{12}\frac{m_a^2}{m_e^2}\right),
    \label{eq:effective_couplings}
\end{align}
where $g_{aN} = m_N/f_a$.

Nevertheless, for light ALPs, $m_a \lesssim 1$ keV, which is the focus of this study, the only open decay channels for the ALPs are either a pair of active neutrinos or a pair of photons. But, as seen in Fig.~\ref{fig:effective_couplings}, the diphoton decay happens at 2-loop and thus it is significantly suppressed with respect to the decay to active neutrinos. For example, the decay width to photons of a 1 keV ALP with decay constant of 1 TeV will be of order $10^{-46}$ GeV, negligible compared to decay width to active neutrinos in eq.~\eqref{eq:anunu}.

The ALP lifetime can thus be computed from its decay width to two active neutrinos. So in ALP rest frame,
\begin{equation}
    \tau_{a} = 1~\rm{sec} \times \left(\frac{1~\rm{GeV}}{m_N}\right)^2 \times \left(\frac{1~\rm{keV}}{m_a}\right) \times \left(\frac{2.03 \times 10^{-6}}{|U_{e N}|^2}\right)^2 \times \left(\frac{f_a}{1~\rm{TeV}}\right)^2.
\end{equation}

For HNLs and ALPs in the mass range of interest, $m_N \sim 1$ GeV and $m_a \sim $ 1 keV, the lifetime of the ALPs, and hence the scenario that is realised, depends on the ALP-HNL interaction $1/f_a$ and the active-sterile mixing $U_{e N}$. For example, an ALP that is stable compared to the age of the universe (i.e., $\tau_a > 10^{17}$ sec) and $f_a \sim 1$ TeV would require $|U_{e N}|^2 \leq 10^{-14.2}$.

\subsection{Benchmark Scenarios}
\begin{table}[t!]
\centering
\renewcommand{\arraystretch}{1.25}
\setlength\tabcolsep{6pt}
\begin{tabular}{c|cccc} 
 \hline
 Scenario & $m_{N}$ [GeV] & $|U_{eN}|^2$ & $f_a$ [TeV]& $m_a$ [keV]\\
 \hline
 1 & $10^{-1}$ & $10^{-10}$ & $1$ & $1$ \\ 
 2 & $10^{-0.4}$ & $10^{-9.2}$ & $10^{2.5}$ & $1$ \\
 3 & - & - & $1$ & $10^{-2}$ \\
 4 & - & - & $10^{2.5}$ & $10^{-2}$\\
 \hline
 \end{tabular}
 \caption{Benchmark scenarios labelled by the red squares on Fig.\ref{fig:parameter_space_Na}. For fixed ALP mass $m_a = 1$ keV, they are $m_N = 10^{-1}$ GeV and $|U_{eN}|^2 = 10^{-10}$ for $f_a = 1$ TeV and $m_N = 10^{-0.4}$ GeV and $|U_{eN}|^2 = 10^{-9.2}$ for $f_a = 10^{2.5}$ TeV respectively. The last two scenarios are given at fixed $m_a = 10$ eV for $f_a = 1$ TeV and $f_a = 10^{2.5}$ TeV, respectively.}
\label{tab:benchmarks}
\end{table}
There are four benchmark points we choose to study are given in Tab.~\ref{tab:benchmarks}. The allowed parameter space is investigated for all four scenarios in active-HNL mixing versus HNL mass planes in Sec.~\ref{sec:results}. The number density evolutions and interaction rates are studied explicitly through cosmological history for scenarios 1 and 2 as examples. These benchmark points satisfy the requirements that HNL decays before the start of BBN and ALP decays before the start of CMB. The reason for these choice is explained in Sec.~\ref{sec:cosmo} and the points are also deliberately chosen in (or close to) the seesaw region in order to explain the masses of active neutrinos.

\section{Cosmological history of HNLs and ALPs}
\label{sec:cosmo}
\begin{figure}[t!]
 \centering
 \begin{tikzpicture}[text centered]
  \draw [fermion]  (-1,1) -> (0,0);
  \node at (-1.5,1.2) {$N$};
  \draw [fermion]  (-1,-1) -> (0,0);
  \node at (-1.5,-1.2) {$N$};
  \draw [fermion]  (0,0) -> (1,1);
  \node at (1.5,1.2) {$SM$};
  \draw [fermion] (0,0) -> (1,-1);
  \node at (1.5,-1.2) {$SM$};
 \end{tikzpicture}
 \begin{tikzpicture}[text centered]
  \draw [fermion]  (-1,1) -> (0,0);
  \node at (-1.5,1.2) {$N/\nu$};
  \draw [fermion]  (-1,-1) -> (0,0);
  \node at (-1.5,-1.2) {$N/\nu$};
  \draw [scalar]  (0,0) -> (1,1);
  \node at (1.5,1.2) {$a$};
  \draw [scalar] (0,0) -> (1,-1);
  \node at (1.5,-1.2) {$a$};
 \end{tikzpicture}
 \begin{tikzpicture}[text centered]
  \draw [fermion]  (-1,0) -> (0,0);
  \node at (-1.5,0) {$N$};
  \draw [scalar]  (0,0) -> (1,1);
  \node at (1.5,1.2) {$a$};
  \draw [fermion] (0,0) -> (1,-1);
  \node at (1.5,-1.2) {$\nu$};
 \end{tikzpicture}
 \begin{tikzpicture}[text centered]
  \draw [scalar]  (-1,0) -> (0,0);
  \node at (-1.5,0) {$a$};
  \draw [fermion]  (0,0) -> (1,1);
  \node at (1.5,1.2) {$\nu$};
  \draw [fermion] (1,-1) -> (0,0);
  \node at (1.5,-1.2) {$\nu$};
 \end{tikzpicture}
 \begin{tikzpicture}[text centered]
  \draw [fermion]  (-1,0) -> (0,0);
  \node at (-1.5,0) {$N$};
  \draw [scalar]  (0,0) -> (1,1);
  \node at (1.5,1.2) {SM};
  \draw [fermion] (0,0) -> (1,-1);
  \node at (1.5,-1.2) {SM};
 \end{tikzpicture}
 \begin{tikzpicture}[text centered]
  \draw [fermion]  (-1,0) -> (0,0);
  \node at (-1.5,0) {$N$};
  \draw [fermion]  (0,0) -> (1,1);
  \node at (1.5,1.2) {SM};
  \draw [fermion]  (0,0) -> (2,0);
  \node at (2.5,0.2) {SM};
  \draw [fermion] (0,0) -> (1,-1);
  \node at (1.5,-1.2) {SM};
 \end{tikzpicture}
 \caption{Relevant processes for the scattering and decay of HNLs, ALPs and SM particles, responsible for determining their abundances. The diagrams on the left describe the scattering of HNLs with SM particles ($NN \leftrightarrow \rm{SM}$), and the scattering of HNLs or active neutrinos with ALPs ($NN \leftrightarrow aa$). The diagrams on the right and second row depict the decays of HNLs ($N \rightarrow a \nu$ and $N \to \rm{SM}$) and ALPs ($a \rightarrow \nu\nu$).}
 \label{fig:relevant_diagrams}
\end{figure}
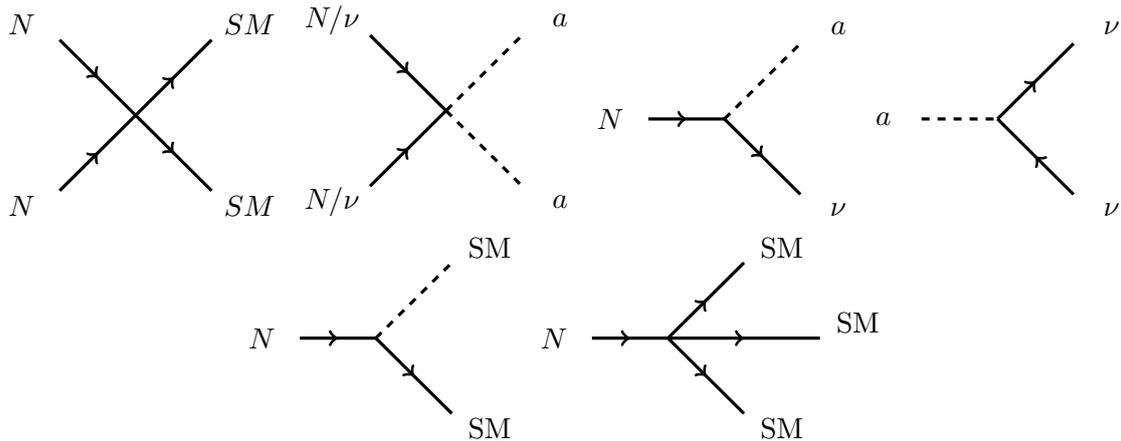

In the hot dense plasma of the early Universe, it is expected that the interactions between the HNLs and SM particles are strong enough to maintain thermal and chemical equilibrium between the two sectors. HNLs with masses around the GeV scale typically do not have time to freeze-out as their decays into SM particles, and in our framework into ALPs, will deplete their abundance sufficiently fast. On the other hand, the interactions of HNLs and SM particles with ALPs are typically weak, and hence we assume that ALPs do not start in thermal equilibrium with the SM and HNLs, and that their initial abundance is negligible (freeze-in production). The scattering between ALPs and HNLs might briefly bring the ALPs in thermal equilibrium, for small enough $f_a$, but they would soon freeze out, and eventually decay into active neutrinos. All of these decay channels and scattering rates play an important role in the evolution of the energy and number densities of HNLs and ALPs throughout the history of the universe. The most relevant processes that determine the abundances of HNLs and ALPs can be seen in Fig.\ref{fig:relevant_diagrams}.

The change on the energy and number densities of the various particle species due to the expansion of the universe and the processes in Fig.~\ref{fig:relevant_diagrams} can be computed with the Boltzmann equations~\cite{Gondolo:1990dk, Escudero:2018mvt}
\begin{align}
    \frac{d\rho_X}{dt} + 3 H(\rho_X+p_X) &= \frac{\delta\rho_X}{\delta t} = \int g_X E\frac{d^3 p}{(2\pi)^3} \mathcal{C}[f], \nonumber\\
    \frac{dn_X}{dt} + 3 H n_X &= \frac{\delta n_X}{\delta t} = \int g_X \frac{d^3 p}{(2\pi)^3} \mathcal{C}[f],
\label{Boltzmann}
\end{align}
where $\rho_X$, $p_X$ and $n_X$ are energy density, pressure and number density of particle $X$, $H$ the Hubble parameter, $g_X$ its internal degrees of freedom and $\delta \rho_X/\delta t$ and $\delta n_X/\delta t$ the energy and number density transfer rates, computed with the collision operator $\mathcal{C}[f]$, which takes into account all energy and number changing processes.

The transfer rates, $\delta\rho_X/\delta t$ and $\delta n_X/\delta t$, can be obtained by calculating the thermally averaged cross sections and decay rates for each process where the species $X$ is involved, $\langle \sigma_{X\to Y} v\rangle$ and $\langle \Gamma_{X\to Y}\rangle$. A useful quantity is the thermally averaged interaction density $\gamma_{X\to Y}$, defined as 
\begin{equation}
 \gamma_{X\rightarrow Y} = \langle \sigma_{X\rightarrow Y} v \rangle n_X^{eq,2} = \langle \Gamma_{X\rightarrow Y} \rangle n_X^{eq},
\label{eq:sigmaGammagamma}
\end{equation}
where $n_X^{eq}$ is the number density of species $X$ while on thermal equilibrium. The thermally averaged decay width density is defined as
\begin{equation}
    \gamma_{X\to Y_1\cdots}= n_X^{eq}(z)\frac{K_1(z)}{K_2(z)}\Gamma_X,
    \label{eq:decay}
\end{equation}
with $z = \frac{m_X}{T}$, $\Gamma_X$ is the zero-temperature decay width of particle $X$, and $K_{1(2)}(z)$ are the Bessel function of first (second) kind. The thermally averaged cross-section density is calculated as,
\begin{equation}
\gamma_{X_1 X_2\leftrightarrow Y_1 Y_2}=\frac{T}{64 \pi^4} \int_{s_{\min }}^{\infty} s^{1 / 2} \hat{\sigma}(s) \mathrm{K}_1\left(\frac{\sqrt{s}}{T}\right) d s,
\label{eq:crosssection}
\end{equation}
where $\hat{\sigma}(s) = 2s \sigma(s)\lambda(1,m_{X_1}^2/s,m_{X_2}^2/s)$ is the reduced cross-section, with the function $\lambda(a,b,c) = (a-b-c)^2-4b c$, and the minimal value of the integral is defined as $s_{min} = \text{Max[(m}_{X_1} + m_{X_2})^2,(m_{Y_1}+m_{Y_2})^2]$.

The addition of the ALPs to this model has the intended consequence of forcing the HNLs to decay faster than in vanilla HNL models. Consequently, even for HNL masses of the order of $10-100$ MeV, HNLs decay fast enough to avoid affecting the BBN abundances. In contrast, the ALPs should not decay before BBN, otherwise they would in turn modify the primordial abundances. Therefore, we study the HNL and ALP abundances in two different time epochs, before BBN and between BBN and recombination\footnote{We should mention that we here consider the onset of BBN to occur at the time of neutrino decoupling, i.e. $t \sim 1$ s. This approximation holds as long as none of the number changing processes involved are active between that time and the end of BBN, $t \sim 10^4$ s.}. After recombination the ALPs may or may not be stable, and that will have some astrophysical consequences that will discussed in Section~\ref{sec:constraints}.

\subsection{Abundances before BBN}
We start by assuming that after some inflationary epoch and subsequent reheating, the SM particles are in equilibrium in the early Universe. The actual value of the reheating temperature is no importance, since most particles will be at their equilibrium densities, but it must be low enough to ensure that the ALPs are not strongly coupled after reheating. The HNLs are thus assumed to be in thermal and kinematical equilibrium with the SM, but the ALPs are not. The evolution of the densities of HNL and ALP can be described by coupled Boltzmann equations, as in eq.~\eqref{Boltzmann}. Before neutrino decoupling, the HNL number density is given by
\begin{equation}
    \begin{aligned}
\frac{d n_N}{d t}+3 H n_N= & -\left\langle\sigma_{N N \rightarrow S M}~ v\right\rangle\left(n_N^2-n_N^{\mathrm{eq}, 2}\right)-\left\langle\sigma_{N N \to a a}~ v\right\rangle\left(n_N^2-n_N^{\mathrm{eq}, 2} \frac{n_a^2}{n_a^{\mathrm{eq}, 2}}\right) \\
& -\left\langle\Gamma_{N \rightarrow S M}\right\rangle\left(n_N-n_N^{\mathrm{eq}}\right)-\left\langle\Gamma_{N \rightarrow a \nu}\right\rangle\left(n_N-n_N^{\mathrm{eq}} \frac{n_a}{n_a^{\mathrm{eq}}}\right),
\end{aligned}
\label{eq:HNLboltz}
\end{equation}
where $\langle\sigma_{NN \rightarrow \rm{SM}}v\rangle$, $\langle\sigma_{NN\to aa}v\rangle$, $\langle\Gamma_{N \rightarrow \rm{SM}}\rangle$ and $\langle\Gamma_{N \rightarrow a \nu}\rangle$ are the thermally-averaged scattering cross-sections of HNL annihilation to SM particles and ALPs, and the thermally-averaged decay widths of HNLs into light SM particles and ALP-active neutrino pair, respectively. Similarly the Boltzmann equation for ALPs is
\begin{equation}
    \begin{aligned}
\frac{d n_a}{d t}+3 H n_a & =-\left\langle\sigma_{a a \rightarrow \nu \nu}~ v\right\rangle\left(n_a^2-n_a^{\mathrm{eq}, 2}\right)-\left\langle\sigma_{a a\to  N N}~ v\right\rangle\left(n_a^2-n_a^{\mathrm{eq}, 2} \frac{n_N^2}{n_N^{\mathrm{eq}, 2}}\right) \\
& -\left\langle\Gamma_{a \rightarrow \nu \nu}\right\rangle\left(n_a-n_a^{\mathrm{eq}}\right)+\left\langle\Gamma_{N \rightarrow a \nu}\right\rangle\left(n_N-n_N^{\mathrm{eq}} \frac{n_a}{n_a^{\mathrm{eq}}}\right),
\end{aligned}
\label{eq:ALPboltz}
\end{equation}
which also contains various contributions from the scattering of ALPs with the HNLs and neutrinos and the thermally averaged ALP decay width $\langle \Gamma_{a\to\nu\nu}\rangle$. The solution of these Boltzmann equations will give the evolution of the number densities of HNLs and ALPs between some unspecified reheating time and the time of neutrino decoupling, $t \sim 1$ s. To solve the Boltzmann equations we make a variable transformation, to the comoving yield $Y_X = n_X/s$ for any species $X$, where $s$ is the entropy density given by
\begin{equation}
  s=\frac{2\pi^2}{45}g_\ast T^3 .
  \label{eq:entropydensity}
\end{equation}
where $g_*$ is the number of relativistic degrees of freedom and $T$ the temperature of the thermal bath of SM particles. The Boltzmann equations for the yield $Y_X$, using the more useful thermally averaged densities $\gamma_{X\to Y}$, can therefore be written as
\begin{align}
z H s \frac{d Y_N}{d z} & = -\gamma_{N N\rightarrow S M}^{\mathrm{eq}}\left(\frac{Y_N^2}{Y_N^{\mathrm{eq}, 2}}-1\right) + \gamma_{a a \rightarrow N N}^{\mathrm{eq}}\left(\frac{Y_a^2}{Y_a^{\mathrm{eq}, 2}} - \frac{Y_N^2}{Y_N^{\mathrm{eq}, 2}}\right) \notag\\
& -\gamma_{N \rightarrow S M}\left(\frac{Y_N}{Y_N^{\mathrm{eq}}}-1\right)-\gamma_{N \rightarrow a \nu}\left(\frac{Y_N}{Y_N^{\mathrm{eq}}}-\frac{Y_a}{Y_a^{\mathrm{eq}}}\right), \\
z H s \frac{d Y_a}{d z} & =-\gamma_{ a a \rightarrow \nu \nu}^{\mathrm{eq}}\left(\frac{Y_a^2}{Y_a^{\mathrm{eq}, 2}}-1\right)-\gamma_{ a a \rightarrow N N}^{\mathrm{eq}}\left(\frac{Y_a^2}{Y_a^{\mathrm{eq}, 2}}-\frac{Y_N^2}{Y_N^{\mathrm{eq}, 2}}\right) \notag\\
& -\gamma_{a \rightarrow \nu \nu}\left(\frac{Y_a}{Y_a^{\mathrm{eq}}}-1\right)+\gamma_{N \rightarrow a \nu}\left(\frac{Y_N}{Y_N^{\mathrm{eq}}}-\frac{Y_a}{Y_a^{\mathrm{eq}}}\right),
\label{eq:HNLALPBEz}
\end{align}
with $z = m_N/T$.

\subsection{Abundance of ALPs after BBN and temperature evolution}
As we saw before, in all scenarios we consider, the abundance of HNLs depletes very fast before BBN. As a consequence, the main production mechanisms for ALPs, which were HNL-ALP scattering and HNL decays, are really inefficient after BBN and thus the ALP abundance only decreases at late times. Even the processes $aa \leftrightarrow N\nu$ and  $aa \leftrightarrow \nu\nu$ fall out of thermal equilibrium significantly before BBN, as seen below in Fig.~\ref{fig:abundance&rates}. Therefore, the abundance of ALPs after the end of BBN is solely determined by their decays. Their Boltzmann equation is therefore given by\cite{Boyarsky:2021yoh}
\begin{equation}
    z H s \frac{d Y_a}{d z} = -\gamma_{a \rightarrow \nu \nu}\left(\frac{Y_a}{Y_a^{\mathrm{eq}}}-1\right), \quad~ \text{with}\quad~z > z_{\rm BBN}.
\end{equation}
If the ALP decays are fast enough, as in the left-hand panel of Figure \ref{fig:abundance&rates}, the ALPs decay shortly after the end of BBN, and their number density completely disappears. Conversely, the abundance of long-lived ALPs becomes frozen-out after BBN, and will only deplete slowly, depending on their lifetime.

Before neutrino decoupling, the temperatures of the photon and neutrino baths evolved together, changing only via adiabatic cooling since ALP decays are negligible before BBN \cite{Hufnagel:2018bjp}. After the end of BBN, however, the decays of ALPs to neutrinos, $a \to \nu\nu$ become relevant, which will introduce energy into the neutrino bath and thus increase its temperature. An increase on the neutrino temperature relative with the photon temperature at the time of recombination has a strong impact on the number of effective neutrino degrees of freedom $N_{\rm eff}$. Because of this, we need to track separately the evolution of the photon and neutrino temperatures from the end of BBN until the formation of the CMB. Neglecting chemical potentials and effects from neutrino oscillations, the evolution of the temperatures with time is given by \cite{Escudero:2018mvt, EscuderoAbenza:2020cmq}
\begin{align}
  \frac{dT_\gamma}{dt} &= - \frac{4 H \rho_\gamma + 3H(\rho_e + p_e)}{\tfrac{\partial \rho_\gamma}{\partial T_\gamma} + \tfrac{\partial \rho_e}{\partial T_\gamma}} \notag \\
  \frac{dT_\nu}{dt} &= - \frac{12 H \rho_\nu + 3 H (\rho_a + p_a) + \delta\rho_a / \delta t}{3\tfrac{\partial \rho_\nu}{\partial T_\nu} + \tfrac{\partial \rho_a}{\partial T_\nu}}
  \label{Tevolution}
\end{align}
where $\rho_i$ and $p_i$ are the energy density and pressure of particle species $i$, and $\delta\rho_a/\delta t$ is the energy exchange rate from ALPs to neutrinos, given by the collision operator in eq.\eqref{Boltzmann}. Equation \eqref{Tevolution} takes into account the decoupling of the electrons from the photon bath when they become non-relativistic, as well as the decoupling of the ALP from the neutrino bath when $T_\nu \gtrsim m_a$.

\subsection{Abundance and interaction rate evolution}
\begin{figure}[t!]
    \centering
    \includegraphics[width=0.49\textwidth]{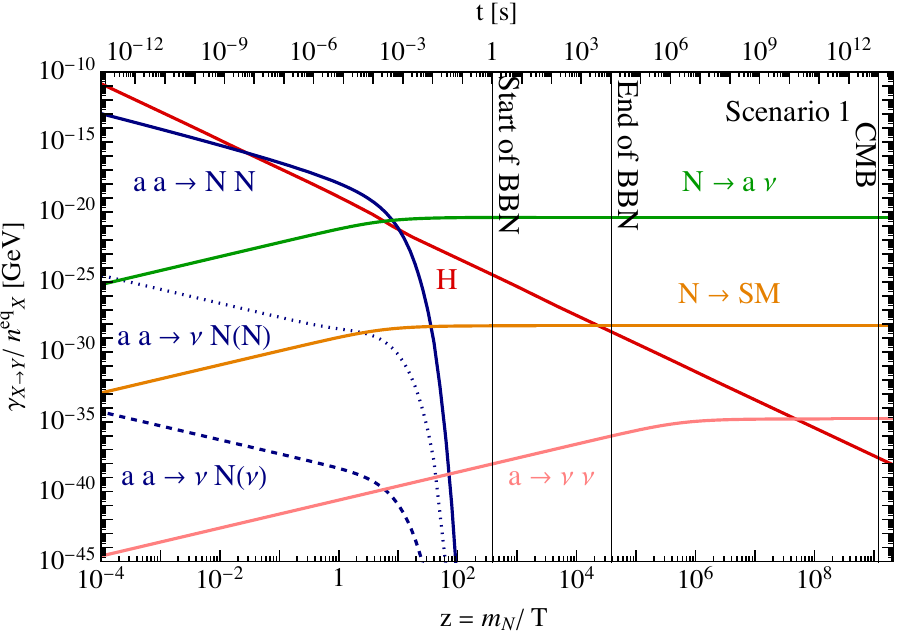}
    \includegraphics[width=0.49\textwidth]{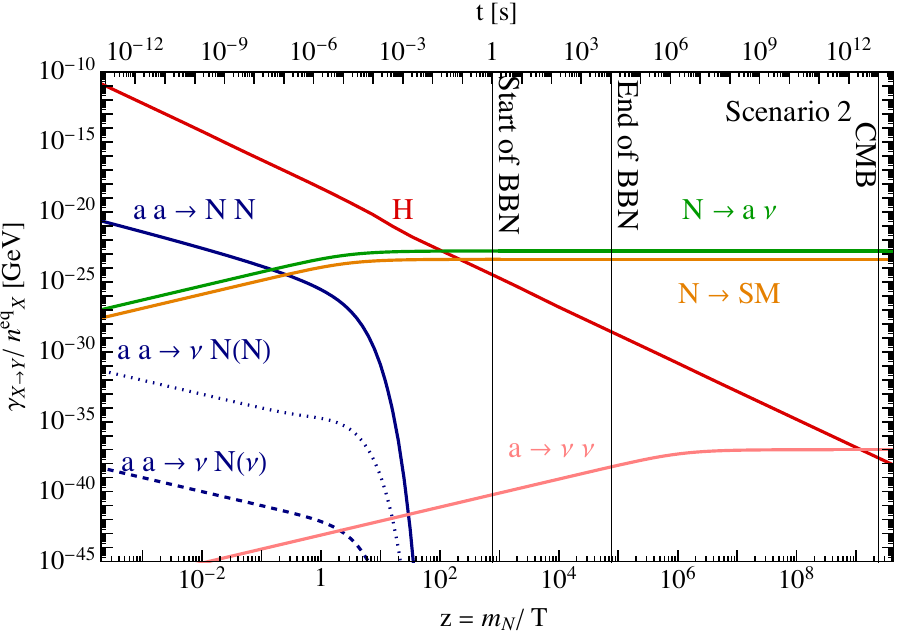}\\
    \includegraphics[width=0.49\textwidth]{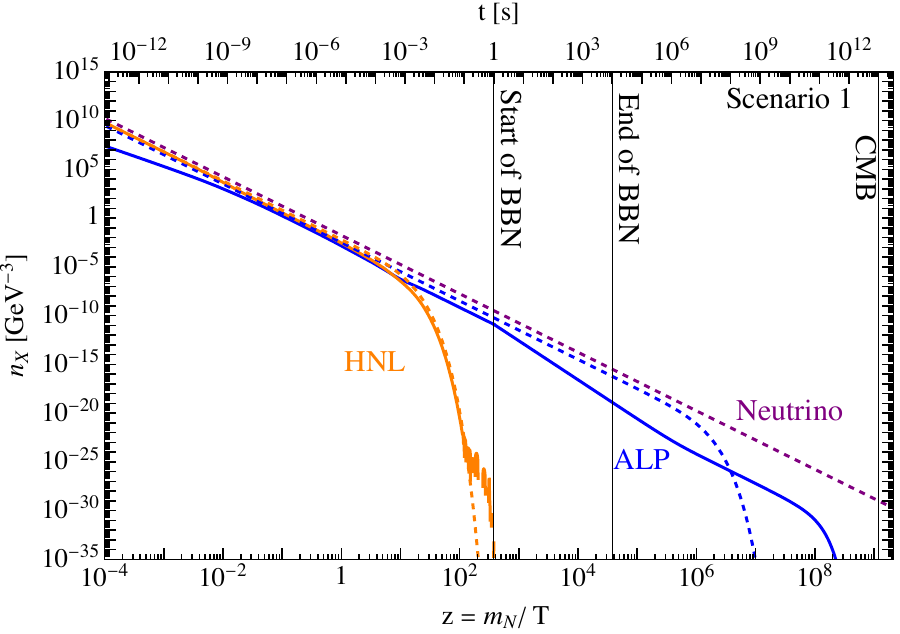}
    \includegraphics[width=0.49\textwidth]{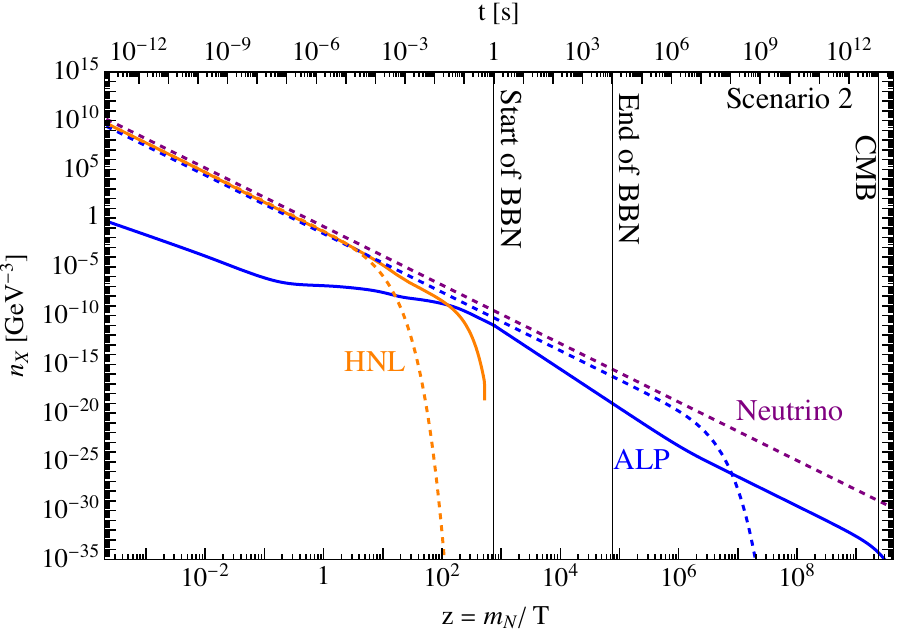}
    \caption{Evolution plots of the thermally averaged scattering and decay rates (top) and number densities (bottom), with $z=m_N/T$ and time, for benchmark scenarios 1 (left) and 2 (right) from Tab.~\ref{tab:benchmarks}. The scattering rates of ALPs with HNLs and neutrinos (blue) are shown in the top row compared to the evolution of the Hubble rate (red), along with the decay rates of HNLs to SM particles (orange) and ALPs (green) and the decay rate of the ALPs (pink). The bottom row shows the number density of ALPs (blue) in and out of equilibrium (dashed and solid respectively), as well as the number densities of HNLs (orange) and neutrinos (purple).}   
    %on the left and the rates for relevant processes on the right for benchmark scenario 1 (left) and 2 (right) as Tab.~\ref{tab:benchmarks} stated. The blue and orange lines represent the number densities for ALP and HNL respectively. The dashed lines denote the equilibrium number densities for the corresponding species and neutrinos (purple line). The Blue lines on the rate plots show the ALP to HNL annihilation rates (solid), ALP to  a neutrino and a HNL annihilation rates via HNL exchange (dotted) and ALP to a neutrino and a HNL annihilation rates via light neutrino exchange (dashed). The green curves give the thermally averaged decay rates for HNL to ALP and a neutrino and the yellow curve give the rates for SM HNL decay channel. There is also the di-neutrino decay rates for ALP which is represented by the pink curves. Finally, the red curves give the evolution of Hubble parameter.}
    \label{fig:abundance&rates}
\end{figure}

Fig.~\ref{fig:abundance&rates} shows the evolution of the thermally averaged scattering and decay rates (top row) and the number density of various species (bottom row) for two different scenarios from Table \ref{tab:benchmarks}. Scenario 1 has $m_N = 10^{-1}$ GeV, $|U_{eN}|^2 = 10^{-10}$, $f_a = 1$ TeV and $m_a = 1$ keV. Scenario 2 has $m_N = 10^{-0.4}$ GeV, $|U_{eN}|^2 = 10^{-9.2}$, $f_a = 10^{2.5}$ TeV and $m_a = 1$ keV.

The rate evolution plots show how the various rates evolve compared to the Hubble parameter (red). In scenario 1, shown on the left, the thermally averaged scattering rate $aa\to NN$ (solid blue) becomes efficient for a short period of time at around $z \sim 1$, which brings the ALPs and HNLs close to thermal equilibrium. Other scattering rates, such as $aa\rightarrow \nu\nu$, with either $N$ (dotted blue) or $\nu$ (dashed blue) mediation, are negligible. Once the ALP-HNL scattering rate becomes inefficient shortly before the onset of BBN, HNL decaying to ALPs (green) become the primary source of ALP production while the HNLs are abundant enough. Other HNL decays to SM particles (orange) only become efficient in scenario 1 much after all HNLs have decayed away to ALPs, and thus have no effect on the evolution. Lastly, a long time after BBN, but before recombination, the decays of ALPs, $a\to \nu\nu$ (pink) become efficient and make the ALPs decay completely before the formation of the CMB. On the other hand, in scenario 2 on the right, the scattering rates are never efficient enough, so the HNLs and ALPs are never in thermal equilibrium. The decay of HNLs to ALPs (green) is thus the only source of ALP production, but in this scenario it is comparable to the HNLs decays to SM particles (orange), and thus not all HNLs decay to ALPs. Since the ALP-neutrino interaction rate is weaker compared to scenario 1, due to larger $f_a$, ALPs do not decay efficiently in scenario 2, as their decay rate only becomes efficient around or after the formation of the CMB, which causes the ALPs to survive after recombination.

In the number density plots in the bottom row, the orange curves show the actual (solid) and equilibrium (dashed) number density for HNL, the blue curves show the actual (solid) and equilibrium (dashed) number density for ALP and the dashed purple curve gives the equilibrium density for neutrinos. In scenario 1, the strong scattering between HNLs and ALPs brings them close to thermal equilibrium at around $z \sim 1$. Shortly after, the HNLs become non-relativistic and thus their actual and equilibrium abundances are Boltzmann suppressed, thereby falling rapidly before BBN. Since the ALPs and HNLs are close to equilibrium already, the increasing effect of HNL decays on the ALP abundance is small, and in fact causes the ALP abundance to decrease as inverse decays tend to dominate when the ALP population is high. Lastly, after BBN the ALPs become non-relativistic, but initially there are no equilibrium processes that can deplete their abundance, and thus they freeze-out, i.e. their real abundance (solid blue) overshoots their equilibrium abundance (dashed blue). Finally, shortly before recombination, ALPs decay to light neutrinos, depleting their abundance completely. In scenario 2, the ALPs are never in thermal equilibrium, and are only efficiently produced by HNL decays, so they freeze-in after the HNLs become non-relativistic and disappear. In this scenario, ALPs are longer lived, and thus their abundance freezes-out, as before, but only depletes slowly before the formation of the CMB.

\section{Constraints from cosmology, astrophysics and direct searches}
\label{sec:constraints}
In this work we have introduced two new particles, a HNL and an ALP that modify very significantly the evolution of the early Universe compared to $\Lambda$CDM. The cosmological and astrophysical implications of both of these species are well known separately~\cite{Cadamuro:2011fd,Marsh:2015xka,Drewes:2015iva}. However, their combination has a somewhat different effect, as the HNLs are allowed to be lighter than in most previous scenarios, due to the weakening of the BBN bound (see below), and also due to the ALPs coupling exclusively to neutrinos via the HNL portal. Consequently we need to evaluate the predictions of our model towards known cosmological and astrophysical observables and hence assess whether their strong constraints render the model invalid or there are still parameter combinations that are unconstrained. For this purpose we study how this model affects the formation of primordial elements (BBN), the observations of the CMB, as well as astrophysical constraints such as those from the observation of the supernova SN1987A, and others such as extra-galactic background light (EBL) or X-ray constraints. Additionally, the new decay channel for HNLs available in this model modifies the prospects of direct searches for HNLs at colliders, fixed target or beam dump experiments. Consequently, we also study the sensitivity of our model to direct searches for HNLs at future facilities.

\subsection{Big Bang Nucleosynthesis}
The predictions of $\Lambda$CDM with regards to the formation of the primordial elements match very well with their observed present abundances\footnote{The present-day abundance of $^7Li$ is currently in disagreement with the predictions from standard cosmology. Attempts have been made to provide an explanation in modified cosmologies, with varying degrees of success~\cite{Depta:2020zbh,Balazs:2022tjl}. Therefore, it remains unclear whether this is caused by modifications over $\Lambda$CDM or an inaccurate measurement of their present abundance.} \cite{Iocco:2008va, Coc:2017pxv}. These predictions are, however very sensitive to the cosmological state of the Universe at any time between neutrino decoupling and the CMB. Any modification of the temperature of neutrino decoupling, the rate of expansion of the Universe or energy injection in the primordial plasma, may cause a catastrophic change on the formation of light elements. This is especially true in our scenario, where the decays of HNLs into mesons disturb the $p\leftrightarrow n$ conversion processes that set the initial proton and neutron abundances for BBN~\cite{Boyarsky:2020dzc}, other works in this direction can be found in \cite{Kohri:2001jx, Kawasaki:2004qu, Kanzaki:2007pd, Kawasaki:2000en, Hasegawa:2019jsa}. If the HNLs decay fast enough, however, the abundances of protons and neutrons have time to restore to the expected values from $\Lambda$CDM, and thus there is no effect on BBN.

\begin{figure}[t!]
    \centering
    \includegraphics[width=0.85\linewidth]{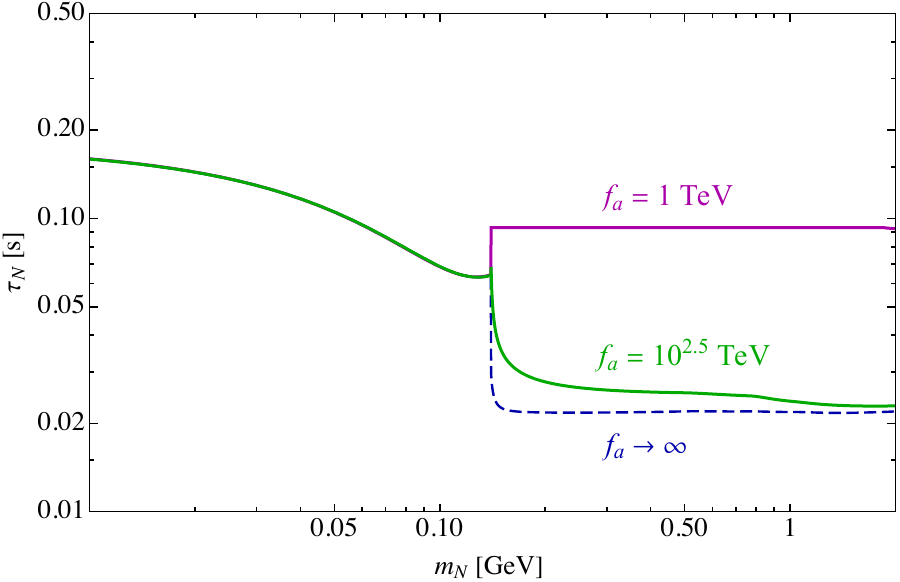}
    \caption{Upper limit on the HNL lifetime $\tau_N$ as a function of its mass $m_N$ by requiring that the HNL decay products do not alter BBN significantly. The plot is adapted from \cite{Boyarsky:2020dzc}, for the case with no ALP (dashed blue) and two scenarios with $f_a = 1$~TeV (magenta) and $f_a = 10^{2.5}$~TeV (green), respectively.}
    \label{fig:HNL-lifetime}
\end{figure}
Such considerations can be used to set an upper limit on the HNL lifetime such that it decays sufficiently early without disturbing the BBN abundances. In Fig.~\ref{fig:HNL-lifetime}, we show the upper limit on the HNL lifetime as a function of its mass. The blue dashed curve corresponds to the typically considered scenario with an HNL only, while the magenta and green solid curves represent scenarios with the ALP and the decay constant $f_a = 1$~TeV and $f_a = 10^{2.5}$~TeV, respectively. The limit is adapted from \cite{Boyarsky:2020dzc} and it is comparable to earlier results in \cite{Dolgov:2000jw, Ruchayskiy:2012si, Sabti:2020yrt}. In the HNL + ALP scenarios, the decay of HNLs to mesons is suppressed, which relaxes the constraint for $m_N \gtrsim m_\pi$.

In addition to the decays of the HNLs, the decays of ALPs may also modify the primordial plasma and affect BBN. If ALPs decay before neutrino decoupling, since their primary decay channel is $a\to\nu\nu$, they would modify the neutrino spectrum and increase the temperature of the neutrino bath, thus delaying neutrino decoupling and the formation of primordial elements. Consequently, we do not consider scenarios where the ALPs have short lifetimes and to ensure that they do not directly affect BBN, we only consider ALPs that decay after the end of BBN, that is, $\tau_a > 10^4$~ s. Subsequent decays of ALPs may also dissociate the formed elements and modify their abundance~\cite{Hufnagel:2018bjp}, however, this dissociation is mostly caused by electromagnetic cascades, which are rare in our scenario, since the decay rate of $a\to \gamma\gamma$ is negligible. In general, late decays of long-lived particles may in principle affect the abundance of light elements through high-energy electromagnetic and hadronic cascades \cite{Kawasaki:1994bs, Kawasaki:2004qu, Kawasaki:2004yh, Kanzaki:2006hm, Jedamzik:2006xz, Ishiwata:2009gs}. Even though our ALP only decays dominantly to neutrinos, cascades and scattering with background neutrinos can dissociate elements, but corresponding constraints apply to heavy long-lived particles whereas we consider light ALPs, $m_a \lesssim 1$~keV that are non-relativistic at the time of decay. The produced neutrinos with kinetic energies $\lesssim 1$~keV are not expected to have any significant impact.

On the other hand, since ALPs are relativistic before neutrino decoupling, a sufficient abundance would act as dark radiation and thus modify the Hubble rate during radiation domination. This effect can be understood as increase in the neutrino temperature and therefore results in a modification on the value of $N_{\rm eff}$ at the time of BBN~\cite{GAMBITCosmologyWorkgroup:2020htv}. Since the contribution from dark radiation is only relevant when the ALPs are close to thermal equilibrium, we can parametrize the deviation from $\Lambda$CDM as
\begin{equation}
    \Delta N_{\rm eff}^{\rm BBN} \approx \frac{\rho_a}{\rho_\gamma} \approx \frac{\rho_a^{eq} n_a}{\rho_\gamma n_a^{eq}}
\end{equation}
where $\rho_\gamma$ is the energy density in photons at BBN. Though $N_{\rm eff}$ is not directly measured at BBN, it can inferred from the measured value at recombination by Planck~\cite{Planck:2018vyg}, to be~\cite{Fields:2019pfx} $N_{\rm eff}^{\rm BBN} = 2.86 \pm 0.15$, hence we set an upper bound on the ALP abundance at BBN by requiring $\Delta N_{\rm eff}^{\rm BBN} \lesssim 0.2$.

\subsection{Cosmic Microwave Background}
After the end of BBN, any particle that injects energy in the primordial plasma will modify, to some extent, the observations of the CMB. The impact of energy injected before recombination would work to heat up the plasma and hence cause changes to the observed anisotropies in the CMB power spectrum. However, since the major source of energy injected is via ALP decays to neutrinos, the energy injected into the photon bath is negligible, and thus we do not expect constraints arising from CMB anisotropies. Late time decays of ALPs could also cause spectral distortions in the black body spectrum of the CMB if the decays heat up the photon bath significantly~\cite{Balazs:2022tjl}. Fortunately, as before, the decay rate from ALPs to photons is negligible and we assume that secondary energy injected into the photon spectrum from the decays of ALPs can be safely ignored.

The most significant impact of the decays of ALPs for recombination is the modification of the neutrino temperature. Since neutrino decoupling, the photon and neutrino temperatures evolved independently which, even in $\Lambda$CDM, causes a value of the effective neutrino degrees of freedom as $N_{\rm eff} = 3.044 \pm 0.384$~\cite{Planck:2018vyg}. ALP decays before the formation of the CMB increase the neutrino temperature, as in eq.\eqref{Tevolution}. This increase of the temperature of the neutrino bath with respect to that of the photon bath, causes the value of $N_{\rm eff}$ to increase from the expectations of $\Lambda$CDM, as
\begin{equation}
    N_{\rm eff}^{\rm CMB} = N_{\rm eff}^{\rm BBN} \left(\frac{11}{4}\right)^{\tfrac{4}{3}}\left(\frac{T_\nu}{T_\gamma}\right)^4
\end{equation}
which strongly constraints the decay rate of ALPs. 

Lastly, from observations of the CMB, the abundance of non-baryonic matter (dark matter) was observed to be around $\Omega_{\rm DM} \sim 0.12$. ALPs that become non-relativistic before recombination, and they have lifetimes larger than $\tau_a > 10^{13}$ s, would survive long enough to contribute their abundance to that of dark matter. Therefore this also sets a strong constraint on the total abundance of ALPs at the time of recombination, which is partially complementary to the constraint on $N_{\rm eff}$.

\subsection{SN1987A}
The core of supernovae (SN) are very hot and dense systems, with temperatures of the order of $T \sim 30 $ MeV. Most of the particles created in such energetic medium are trapped and contribute to the energy transfer inside the SN core. Weakly coupled particles, however, can free-stream and escape the core, contributing to the cooling of the SN. The primary source of cooling for SNs are neutrinos, which can escape as long as they have energies $E_\nu \lesssim 30$ MeV. This neutrino burst was observed for SN1987A by various water Cherenkov detectors, including Kamiokande-II~\cite{Kamiokande-II:1987idp, Totsuka:1988iyh}. Besides neutrinos, the HNLs and ALPs in our model could also escape the SN core as they interact very weakly with the SN inner medium, and contribute to SN cooling~\cite{Dolgov:2000pj, Fuller:2008erj, Mastrototaro:2019vug, Fiorillo:2022cdq, Syvolap:2023trc, Akita:2023iwq}. This additional source of cooling is very constrained by measurements of the luminosity of SN1987A. Secondary decays of HNLs and ALPs into neutrinos produce a high-energetic additional flux of active neutrinos that could have been detected alongside the normal neutrino burst~\cite{Fiorillo:2022cdq, Syvolap:2023trc}. The combination of SN cooling and the additional neutrino flux can impose strong constraints on our model for high couplings.

For both HNLs and ALPs, the constraint on the secondary neutrino flux is much stronger than the constraint from SN cooling~\cite{Syvolap:2023trc, Akita:2023iwq}, and hence we only include the former in our study. Typical constraints on the secondary neutrino flux from HNL decays assume decay rates for HNLs in the absence of additional channels, so the $N\to \pi\nu$ and $N\to 3\nu$ decays dominates~\cite{Mastrototaro:2019vug}. Other studies involving SN constraints on lighter HNLs can be found in \cite{Kainulainen:1990bn, Raffelt:1992bs, Dolgov:2000pj, Telalovic:2024cot}. In our model the primary decay channel for HNLs is to ALPs, which consecutively decay to a pair of neutrinos. Hence, neglecting resonant effects, the secondary neutrino flux from HNLs can be approximated to that arising from the 3-body decay of the HNLs to neutrinos. Even though the 2-body decay branching ratio is around 7 times larger than the 3-body decays for HNL masses in the 100 MeV range, Cherenkov detectors are more sensitive to the 3-body decays, since the interaction cross section of antineutrinos, only produced in 3-body decays, is around 100 times larger than that of neutrinos~\cite{Kamiokande-II:1987idp}. Therefore, SN limits on traditional HNL decays can be directly applied to our scenario. We thus use the limits on HNL masses and mixing from Fig.~2 of \cite{Akita:2023iwq} to constrain the large mixing angle regions in our model.

In addition to HNLs, ALPs can also be produced in the core of SN. The expected secondary neutrino flux from decaying ALPs with masses below the keV scale is smaller than that from HNL decays that can escape the SN core. However, heavy HNLs cannot escape and thus SN constraints from the production of ALPs are stronger for $m_N \gtrsim 400$ MeV. From Fig.~5 of \cite{Akita:2023iwq}, we can see that the coupling between the ALP and the electron-neutrino $g_{ae}$ must be $g_{ae} \lesssim 10^{-7}$ for keV-scale ALPs. In our model, $g_{ae}$ is given by Eq.~\eqref{eq:effective_couplings}, which imposes only weak limits for heavy HNLs and large HNL-ALP couplings, beyond the ranges considered in our study.

Lastly, there could be additional constraints from secondary decays of HNLs and/or ALPs to photons~\cite{Jaeckel:2017tud, Hoof:2022xbe, Diamond:2023scc, Diamond:2023cto, Caputo:2022mah, Caputo:2021rux}. However, these are negligible as the photonic branching ratios of both particles are very small for the masses considered.

\subsection{Other astrophysical constraints}
In addition to the astrophysical constraints from SN cooling and its secondary fluxes, there is a plethora of possible astrophysical probes of ALPs. The production of ALP in the core of white dwarfs or RGB stars can lead to strong cooling, and even provide an explanation for the observed cooling hints~\cite{Giannotti:2015kwo,DiLuzio:2020wdo}. In fact, it has been shown that an ALP with a coupling to electrons of the order $g_{ae} \sim 10^{-13}$ provides a good fit to stellar cooling data~\cite{DiLuzio:2020wdo}. However, in our model the ALP-electron coupling is very small, so much that the value $g_{ae} \sim 10^{-13}$ can only be reached for mixing $|U_{eN}|^2 \sim 10^{-1}$ and HNL masses $m_N \sim 1$ TeV. Since this parameter region is beyond our scope, and anyways disfavoured by various cosmological constraints, as we will see later, we neglect the constraints from stellar cooling in our study.

Long-lived ALPs that survive after recombination may still be observable today through their decay products. The photons injected by ALP decays are the most detectable candidates, as they can be probed in observations of the extra-galactic background light (EBL) or via X-rays~\cite{Cadamuro:2011fd}. For ALP masses below the keV scale, the constraints on the ALP-photon coupling from X-ray and EBL surveys require that $g_{a\gamma} \lesssim 10^{-15}$ GeV$^{-1}$~\cite{Cadamuro:2011fd, Balazs:2022tjl}. In our model, however, the ALP-photon coupling is derived at two-loop order, and therefore extremely small. As above, only for very large mixing $U_{eN}^2 \sim 10^{-3}$ and large HNL masses $m_N \sim 1$ TeV, has this constraint any effect. Consequently, we also ignore any astrophysical constraints from late-time photonic decays of ALPs.

\subsection{Impact on direct HNL searches}
\label{sec:direct-searches}
Future HNL searches, especially those based on long-lived signatures, will probe small active-sterile mixing strengths approaching the seesaw expectation. Prominent examples of proposed and planned searches are PIONEER \cite{PIONEER:2022yag}, NA62 \cite{Dias:2022fwf, Drewes:2018gkc}, DUNE \cite{Ballett:2019bgd}, SHiP \cite{SHiP:2018xqw}, FCC-ee \cite{Blondel:2022qqo}. A dedicated analysis is needed to present a study of the sensitivity in our scenario depending on the specific search strategy. Such a detailed analysis is beyond the scope of this paper and will be considered as future work. In this section, we will assess how the sensitivity is modified in presence of new HNL-ALP coupling in the context of DUNE. We qualitatively describe the approach to calculate the expected number of signal events in the DUNE near detector (ND) based on the analysis \cite{Bolton:2022tds}. The ND is located at a distance of $L = 574$~m from the HNL production point, with a transverse cross-section of $A = 12~\rm{m}^2$ and a depth of $\Delta L = 5$~m along the beam axis. Following \cite{Bolton:2022tds}, we write the expected number of signal events as
\begin{equation}
    N_{\rm{sig}} = N_P \times \rm{Br}(P \rightarrow N) \times \rm{Br}(N \rightarrow \rm{charged}) \times \epsilon_{\rm{geo}},
\end{equation}
where $N_P$ is the relevant production fraction of positively-charged and neutral pseudoscalar mesons multiplied by the total number of protons on target $N_{\rm{POT}} = 6.6 \times 10^{21}$ for a 120 GeV proton beam at DUNE, $\rm{Br(P \rightarrow N)}$ is the branching fractions of HNL production from the meson $P$, $\rm{Br} (N \rightarrow \rm{charged})$ is the branching fraction of produced HNL decaying into charged lepton pairs and $\epsilon_{\rm{geo}}$ corresponds to the geometrical efficiency,
\begin{equation}
    \epsilon_{\rm{geo}} = e^{-\frac{m_N \Gamma_N}{p_{N_z}}L} 
    \left(1-e^{-\frac{m_N \Gamma_N}{p_{N_z}}\Delta L}\right).
\end{equation}
Here, $p_{N_z}$ is the momentum of the HNL along the beam axis in the lab frame, where we have considered $p_{N_z} = 7.5$~GeV, following the simulation of
meson production from a $pp$ collision at $\sqrt{s} = 15$~GeV \cite{Bolton:2022tds}. As we have not explicitly simulated any events and analysed with respect to detector level cuts, for simplicity we will consider, in our analysis, that all produced HNL events will be accepted at detector level.

In presence of the HNL-ALP coupling as well as the new HNL decay channel $N \rightarrow a \nu$ in our scenario, the corresponding total decay width will also be affected. Correspondingly, the partial decay width of HNLs decaying into visible final states as well as the geometrical efficiency factor will change. In contrast, the production rate $N_P$ as well as the branching ratio of HNL production from mesons is unaffected. We thus consider the relative change in the number of events,
\begin{equation}
\label{eq:Nsigratio}
    \frac{N^\prime_{\rm{sig}}}{N_{\rm{sig}}} = 
    \frac{\rm{Br^\prime}(N \rightarrow \rm{charged})}{\rm{Br}(N \rightarrow \rm{charged})}
    \frac{\epsilon^\prime_{\rm{geo}}}{\epsilon_{\rm{geo}}}
\end{equation}
where the unprimed and primed quantities correspond to $\Gamma_N \equiv \Gamma^{N \rightarrow \rm{SM}}$ and $\Gamma^\prime_N \equiv \Gamma^{N \rightarrow \rm{SM}} + \Gamma^{N \rightarrow a \nu}$, respectively, as discussed in Sec.~\ref{subsec:HNL_decays}. From Fig.~\ref{fig:brN}, we can notice that for $(m_a,f_a) = (1~\rm{keV}, 1~\rm{TeV})$, the total decay width of the HNL is dominated by $N \rightarrow a \nu$ throughout the considered mass range, while for ($1~\rm{keV}, 10^{2.5}~\rm{TeV}$), up to $m_N \sim m_\pi$, $N \rightarrow a \nu$ dominates the scenario and beyond this point, it drops significantly. Furthermore, as the decay widths for $N \rightarrow a \nu$ and $N \rightarrow \rm{SM}$ have the same $U_{eN}$ dependence, the branching fractions are independent of $U_{eN}$.

Likewise, the ratio of the geometric efficiencies can be approximated by 
\begin{align}
    \frac{\epsilon^\prime_{\rm{geo}}}{\epsilon_{\rm{geo}}} = 
    \exp\left[-\frac{m_N}{p_{N_z}} \Gamma(N\to a\nu) L\right]
    \frac{\Gamma^\prime_N}{\Gamma_N},
    \label{eq:repsilon}
\end{align}
for a shallow detector depth $\Delta L$, i.e., for $(m_N/p_{N_z})\Gamma'_N \Delta L \ll 1$. The ratio of decay rates in this expression is equal to $\Gamma^\prime_N / \Gamma_N = \text{Br}(N\to\text{charged}) / \rm{Br'}(N\to\text{charged})$, thus cancelling the corresponding ratio in Eq.~\eqref{eq:Nsigratio} in this limit. We thus have 
\begin{equation}
\label{eq:Nsigratio2}
    \frac{N^\prime_{\rm{sig}}}{N_{\rm{sig}}} = 
    \exp\left[-\frac{m_N}{p_{N_z}} \Gamma(N\to a\nu) L\right],
\end{equation}
which approaches unity for long decay lengths, $(m_N / p_{N_z}) \Gamma(N\to a\nu) L \ll 1$. Thus, for small $|U_{eN}|^2$ and $m_N$ where $\Gamma(N\to a\nu) \ll p_{N_z}/(m_N L)$, we expect that DUNE will have the same sensitivity towards the active-sterile mixing strength $|U_{eN}|^2$ in our HNL-ALP scenario as in the standard HNL case. For shorter decay lengths, the sensitivity will be reduced.

As another example for a direct search experiment we consider the existing NA62 experiment which uses a different search strategy as compared to DUNE. The NA62 experiment \cite{Dias:2022fwf} used a secondary 75~GeV hadron beam containing a
fraction of kaons, and has been able to probe the decays $K^+ \to \ell^+ N$. For small active-sterile mixing the decay length of HNL is much larger than the 75~m detector size and the process is characterised by a single detected track, that of the charged lepton – a positive signal is a peak in its missing mass distribution. As this experiment is currently insensitive to the decay of the HNLs, the presence of the $N \to a\nu$ decay channel will not affect the search as long as the decay length remains large.

\section{Results and discussion}
\label{sec:results}
\begin{figure}[t!]
   \centering
   \includegraphics[width=0.49\textwidth]{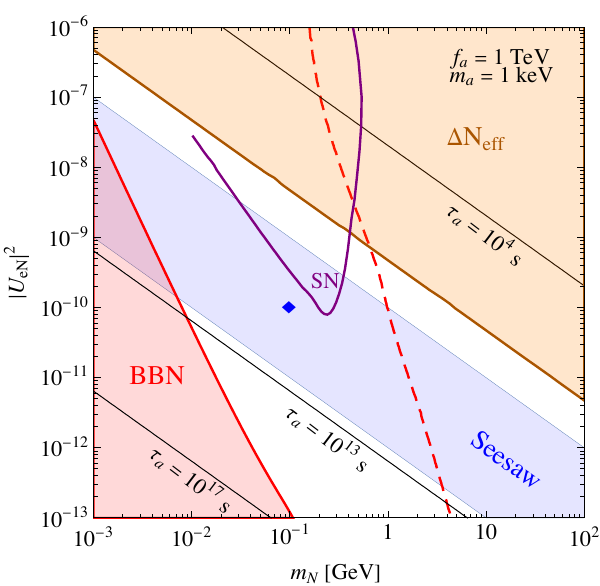}    \includegraphics[width=0.49\textwidth]{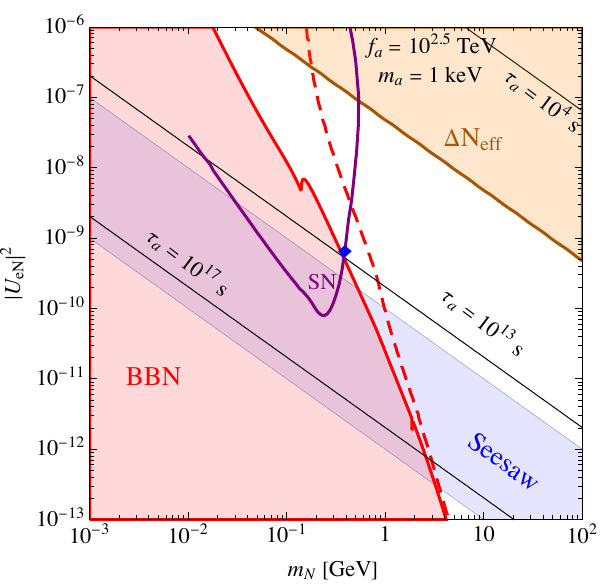}\\
   \includegraphics[width=0.49\textwidth]{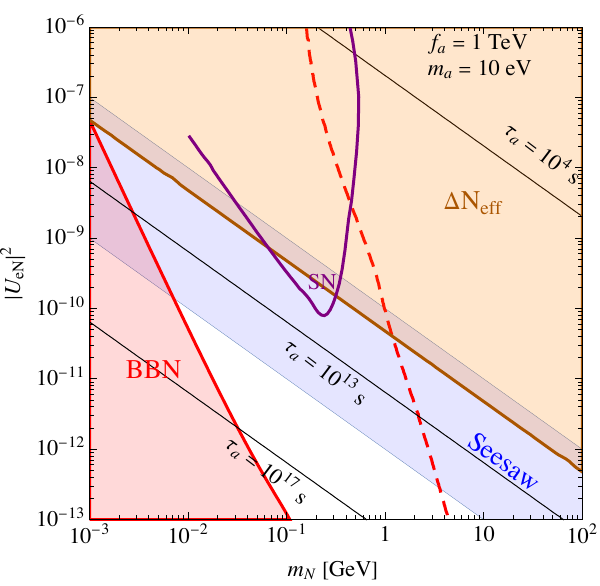}
   \includegraphics[width=0.49\textwidth]{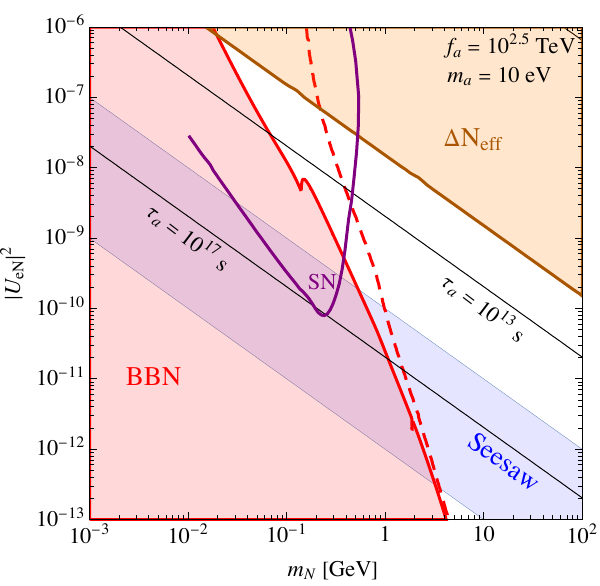}
   \caption{Allowed parameter regions in the $m_N$ vs $|U_{eN}|^2$ plane for the four benchmark scenarios in Table \ref{tab:benchmarks}: Upper left panel : $f_a = 1$ TeV, $m_a = 1$ keV, Upper right panel : $f_a = 10^{2.5}$ TeV, $m_a = 1$ keV, Lower left panel : $f_a = 1$ TeV, $m_a = 10$ eV, Lower right panel : $f_a = 10^{2.5}$ TeV, $m_a = 10$ eV. The red region shows the space disallowed by HNL decay time after the start of BBN, while the region left to red dashed contour is disfavored if HNL decays only via SM decay channels, in absence of ALP. The most stringent bound from cosmology, $N_\text{eff} < 3.10322~ (\Lambda\text{CDM}) + 0.0384~ (0.1\sigma_N)$ gives the brown forbidden region of active neutrinos produced from ALP decays. The region above the purple contour is excluded due to astrophysical constraints from supernova SN1987A. The seesaw regime for active neutrinos $ 1~\rm{meV}~\lesssim m_\nu \lesssim 100~\rm{meV}$ is shown in blue. The different black lines label the ALP decay time at $10^{4}$ s (end of BBN), $10^{13}$ s (CMB) and $10^{17}$ s (today). Benchmark values from Tab.\ref{tab:benchmarks} for scenario 1 and 2 with fixed $m_N$ and $|U_{eN}|^2$ are denoted in top panels by blue diamonds.}
\label{fig:parameter_space_Na}
\end{figure}

The aim of this work is to find valid scenarios where the traditional BBN bound on HNLs \cite{Boyarsky:2020dzc} is relaxed due to the primary decays to ALPs. For this purpose we have performed some small scale parameter scans around the benchmark scenarios in Table \ref{tab:benchmarks} to illustrate this effect\footnote{We performed a grid scan on two parameters of the model, $m_N$ and $|U_{eN}|^2$. This simplification is enough for our purposes as we only intend to highlight the crucial features of the model, but do not attempt to map the available parameter space nor perform any statistical interpretation of the results. For more details on the rigorous treatment of inference and statistics in physics see \cite{AbdusSalam:2020rdj}.}. Figure \ref{fig:parameter_space_Na} shows the results of these parameter scans, varying the values of $m_N = [10^{-3},10^2]$ GeV and $|U_{eN}|^2 = [10^{-13},10^{-6}]$. The top row corresponds to scans around scenarios 1 and 2, with only $m_a$ and $f_a$ fixed, and, where the blue diamond markers are the fixed values of $m_N$ and $|U_{eN}|^2$ from Table \ref{tab:benchmarks}. The bottom row shows scans of scenarios 3 and 4, which have a lighter ALP mass of $m_a = 10$ eV.

In all the panels of Fig.~\ref{fig:parameter_space_Na}, the three different regimes of ALP decay lifetime (as discussed earlier) are shown. ALPs decaying before the end of BBN, i.e., $\tau_a < 10^4$ s, to the right of the $\tau_a = 10^4$ s line, are not considered in this study. As we will argue below, points in this region are excluded anyway due to the fast ALP decays, so in this way we are justified in neglecting scenarios with short-lived ALPs. As expected, since the lifetime of the ALP goes like $\sim f_a^2/m_a$, for the larger values of $f_a$ in the right-hand panels and lower $m_a$ values on the bottom panels, the excluded region for short ALP lifetimes moves to larger HNL masses and larger mixing. Additionally, the ALP lifetimes around the formation of the CMB, $\tau_a = 10^{13}$ s, and the age of the Universe, $\tau_a = 10^{17}$ s are shown.

The dashed red line corresponds to the HNL decay lifetime $\tau_N = 0.023$ s where only SM decay channels are present, i.e. in the absence of the ALP. Conversely, the solid red line denotes the same HNL lifetime limit, but in the presence of the additional decay channel $N \rightarrow a \nu$. Since HNL decays after the start of BBN would modify the abundance of primordial elements, scenarios with longer lived HNLs are also not considered, and hence shaded in red in Fig.~\ref{fig:parameter_space_Na}. Consequently, whenever the HNL to ALP decays dominate, the BBN limit is relaxed as compared to the standard scenario. This effect is more evident on the left-hand side panels, with $f_a = 1$ TeV, where the BBN limit is lowered by about two orders of magnitude with respect to vanilla HNL models. For larger $f_a$ values, on the right-hand panels, the HNL-ALP coupling, which goes like $\sim 1/f_a$, is weaker and thus the exclusion due to the HNL lifetime is stronger as the impact of the addition of the ALP to the model is less significant.

In this article, for simplicity, we have only considered a single HNL, which is not enough to generate non-zero masses for all active neutrinos. Nevertheless, it is useful to show the expected mass scale of active neutrino mass generation, under the seesaw approximation. We thus show in Fig.~\ref{fig:parameter_space_Na} the blue shaded region for the seesaw mass $1~\rm{meV}~ \lesssim m_\nu \lesssim 100$ meV. 

 The strongest cosmological constraint, shown as the shaded brown region in the panels, arises from the increase on the neutrino temperature due to ALP decays, which shows as a modification of the $N_{\rm eff}$ at the time of recombination. $N_{\rm eff}$ constrains the larger HNL masses and mixing, corresponding to shorter ALP lifetimes. Consequently, ALPs decaying before the onset of BBN, or shortly after, would cause a sufficient increase of the neutrino temperature and be excluded by the Planck measurement of the $N_{\rm eff}$ value. Smaller HNL-ALP couplings cause longer ALP lifetimes, and thus the right-handed panels with larger $f_a$ have weaker $N_{\rm eff}$ constraints. However, for the bottom panels with lower $m_a$, even though the lifetime is also longer and one would expect the limit to be weaken, the opposite effect occurs and the $N_{\rm eff}$ limit is significantly stronger. This happens because lighter ALPs are ultra-relativistic for a longer period between BBN and recombination, and thus have a stronger effect on the expansion rate, and thus on the neutrino temperature and value of $N_{\rm eff}$. Other cosmological constraints such as, energy injection before BBN or the relic abundance of ALPs are subleading and thus not shown. The region above the purple contour is, in principle, excluded due to the astrophysical constraints from supernova SN1987A, arising from the modified neutrino flux due to the production of HNLs in the core of the supernova. Nevertheless, the uncertainties in the calculation of the SN neutrino flux are high and depend on the choice of SN core model~\cite{Carenza:2023old}, hence we show in purple a conservative expectation of the limit, but refrain from excluding models only on the basis of this SN constraint. Supernova constraints due to ALP production, which is largely independent of $m_a$ in the ranges of interest, only apply for large values of $|U_{eN}|^2$ and $m_N$, which however resides well outside of our considered region. 

The combination of all the overlaid constraints still leaves a significant region of the parameter space allowed, which becomes narrower for large values of $|U_{eN}|^2$ and wider for smaller values. The most interesting section of this allowed region corresponds to the one on the left of the BBN limit in vanilla HNL models (dashed red), which is the parameter space gained for lower HNL masses with the introduction of the additional decay channel $N\to a \nu$. The panels on the left, for $f_a = 1$ TeV, show a significant increase on the viability of HNL masses, all the way down to $m_N \sim $~MeV, at the expense of stronger constraints on the mixing $|U_{eN}|^2$ due to the stronger HNL-ALP interactions. The newly open region is much narrower in the panels on the right, with $f_a = 10^{2.5}$ TeV, as the HNL-ALP interactions are much weaker, and only slightly relevant for high $|U_{eN}|^2$ values. On the other hand, stronger HNL-ALP interactions, for $f_a < 1$ TeV, would not provide any noticeable improvement, as the ALPs would then be produced in equilibrium and the cosmological constraints would be much stronger. In conclusion, from Fig.~\ref{fig:parameter_space_Na} there is a reasonable expectation for HNLs in our model to have masses at around the MeV scale without affecting the cosmological history of the Universe.

\begin{figure}[t!]
    \centering
    \includegraphics[width=0.49\linewidth]{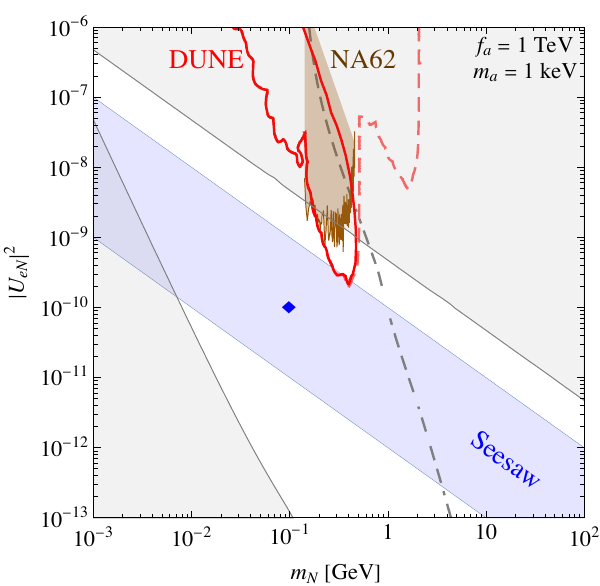}
    \includegraphics[width=0.49\linewidth]{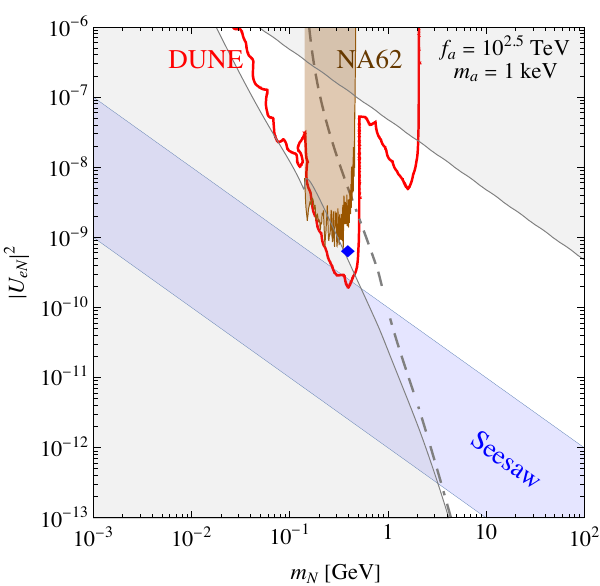}
    \caption{The sensitivity contours of future DUNE and current NA62 searches for observation of 6 events in newly available parameter space constrained by several astrophysical and cosmological constraints (gray shaded region) for two different benchmark scenarios i.e., left panel ($f_a$ = 1 TeV, $m_a = 1$ keV) and right panel ($f_a = 10^{2.5}$ TeV, $m_a = 1$ keV). The region left to gray dashed contour is disfavored if HNL decays only via SM decay channels, in absence of ALP. The seesaw regime for active neutrinos $ 1~\rm{meV}~\lesssim m_\nu \lesssim 100~\rm{meV}$ is shown in blue. Benchmark values from Tab.\ref{tab:benchmarks} for scenario 1 and 2 with fixed $m_N$ and $|U_{eN}|^2$ are denoted in two panels by blue diamonds. The red solid (dashed) contours denote the DUNE sensitivity contours with (without) HNL to ALP decay channel. The brown contour corresponds to NA62 present sensitivity while the brown shaded region denotes the unaffected parameter space, in presence of $N \rightarrow a \nu$ channel, using the approximation discussed in subsection \ref{sec:direct-searches}.}
    \label{fig:dune}
\end{figure}

In Fig.~\ref{fig:dune} we present this newly available parameter region which may be the target of future dedicated searches for HNLs. In the left panel, we show the corresponding parameter space for scenario 1 ($f_a = 1$ TeV, $m_a = 1 $ keV) as described in previous section, while the right panel corresponds to the scenario 2 ($f_a = 10^{2.5}$ TeV, $m_a = 1$ keV). The gray shaded regions in both panels are disfavored from various cosmological and astrophysical constraints discussed in Sec. \ref{sec:constraints}, and shown coloured in Figure \ref{fig:parameter_space_Na}. The gray dashed curve denotes the BBN bound from the HNL only decaying to SM decay channels i.e., in absence of the ALP. The corresponding seesaw regime is shown in blue considering $1~\rm{meV}~ \lesssim m_\nu \lesssim 100~\rm{meV}$. The red dashed contour corresponds to the future sensitivity of DUNE HNL searches in standard HNL models (without ALPs), while the solid red contours in both panels denote the resultant sensitivity contours for observation of 6 events in presence of the new $N \rightarrow a \nu$ decay channel in respective benchmark scenarios with two different $f_a$ values. Additionally, as described in Tab.~\ref{tab:benchmarks}, in blue diamonds two benchmark scenarios with fixed HNL mass and mixing $m_N = 10^{-1}$ GeV, $|U_{eN}|^2 = 10^{-10}$ and $m_N = 10^{-0.4}$ GeV, $|U_{eN}|^2 = 10^{-9.2}$, have been shown. The brown dashed contour denotes the present exclusion contour from NA62 experiment. Furthermore, considering the discussion of subsection \ref{sec:direct-searches}, we have also highlighted in brown the region where the approximation holds (with $L_{\rm{NA}62}$ = 75 m and $p_{N_z} \sim$ 30 GeV as the $K^+$ beams have energy $\sim 75$ GeV \cite{Dias:2022fwf}), the present sensitivity contour remains unaffected by the new axion decay channel of HNL in this region. From Figure \ref{fig:dune} it is evident that the available regions constrained by several astrophysical and cosmological constraints can be well probed in the future by DUNE and NA62 in presence of this new decay channel of HNLs.

\section{Conclusions and outlook}
\label{sec:conclusions}
In this work we have presented a simplified model of new physics including a single HNL, which mixes with a light active neutrino in a seesaw scenario, and an axion-like particle, whose only tree-level interaction is with the HNL. This model is motivated to explore the potential impact on the decay rate of the HNL allowing to evade constraints from BBN for HNL masses around the MeV scale. This would open up parameter space probed by HNL direct searches that is otherwise disfavoured by cosmological considerations. However, the ALPs produced by the scattering and decay of HNLs can have a significant impact on the evolution of the Universe, as they affect its expansion rate and their own decays increase the temperature of the neutrino bath. Further constraints on the model arise from the production of ALPs in the core of stars or supernovae, which is strongly constrained from the observation of the photon and neutrino fluxes from SN1987A.

We have thus found that, in spite of the strong constraints, there is viable parameter space in the region of interest where the ALPs are not produced abundantly enough to modify the cosmology of the Universe significantly. They nevertheless open up a new and dominant decay channel for the HNLs so that they decay non-hadronically and are not constrained by BBN. The relaxation of the BBN limit compared to the standard HNL model is more significant for stronger HNL-ALP interactions ($f_a = 1$ TeV), where HNL masses around $m_N \sim $~MeV are allowed, with a relatively large active-sterile mixing. 

This region of parameter space with HNL masses $1~\text{MeV} < m_N < 1$~GeV and couplings $10^{-10} < |U_{eN}|^2 < 10^{-7}$, which is allowed in our model in contrast to the vanilla HNL model, is a prime target for direct searches for HNLs at colliders, fixed target and beam dump experiments. While a detailed analysis is beyond the scope of this work, we discuss the constraints from the existing NA62 search and the expected sensitivity of a future search at the DUNE near detector. We find that the additional decay to ALPs significantly alters the probed parameter space but as long the HNL remains sufficiently long-lived, i.e., for $m_N \approx 100$~MeV to 1~GeV, both experiments still probe small active-sterile mixing strengths $|U_{eN}|^2 \approx 10^{-9}$.

Our simplified model motivates the search for lighter HNLs within a region otherwise disfavoured by BBN constraints. It also shows that, once allowing the HNL to couple to a light dark sector, the phenomenology, specifically as a long-lived particle can change significantly, and searches should consider non-standard decay lengths and exotic decay modes. This is in correlation with cosmological observations and future improvements in this sector are expected to tighten the presence of a light sector.

\acknowledgments

The authors would like to thank Patrick D. Bolton, Torsten Bringmann, Felix Kahlhoefer, Mudit Rai and Robert Ziegler for useful suggestions and discussions. TEG acknowledges funding by the Deutsche Forschungsgemeinschaft (DFG) through the Emmy Noether Grant No. KA 4662/1-2. CM acknowledges the support from the Royal Society, UK, through the Newton International Fellowship (grant number NIF$\backslash$R1$\backslash$221737). FFD acknowledges support from the U.K. Science and Technology Facilities Council (STFC) via the Consolidated Grant ST/X000613/1.

\appendix

\section{Calculation of decay and scattering rates}
In this appendix, we will give analytical expressions for the scattering cross-sections involving active neutrinos, HNL and ALP in all possible way. Due to the couplings presented in Eq.\eqref{eq:Lsimplified} one can also consider the axion-HNL scattering in early universe. The process happens via \textit{t}-channel scattering. The matrix element responsible for the process $a (p_1) + a (p_2) \rightarrow N (p_3,s) + N (p_4,r)$ can be written as
\begin{equation}
    \mathcal{M}_t = \frac{4m_N^2}{f_a^2} \left(\bar{u}^s_3 \frac{\slashed{q}-m_N}{q^2 - m_N^2} u^r_4\right)
\end{equation}
where momentum exchange is $q^2 \equiv (p_1-p_3)^2 = t$. Now finding $|\mathcal{M}_t|^2$ and performing the sum over final spin states $s,r$ of HNLs one gets
\begin{align}
    \sum_{r,s} |\mathcal{M}_t|^2 = \frac{64m_N^4}{f_a^4 (t-m_N^2)^2}&\Big[2(p_3.q)(p_4.q)-(p_3.p_4)(q.q-m_N^2)+2m_N^2(p_3.q)+2m_N^2(p_4.q)\nonumber \\
    &+m_N^2(q.q+m_N^2)\Big]
\end{align}
Now, putting the definitions for kinematic variables and phase space factor as well as integrating over the angular variables, we get the total cross-section as
\begin{align}
    \sigma_{aaNN} = \frac{4m_N^4 \sqrt{s-4m_a^2}}{\pi f_a^4 s^{3/2}}
    \Bigg[ &-1-\frac{m_a^4-4m_a^2m_N^2}{m_a^4-4m_a^2m_N^2+m_N^2 s} \nonumber\\
    &+ \frac{2(s-2m_a^2)\rm{coth}^{-1}\left(\frac{s-2m_a^2}{\sqrt{s-4m_a^2}\sqrt{s-4m_N^2}}\right)}{\sqrt{s-4m_a^2}\sqrt{s-4m_N^2}}\Bigg].
\end{align}
Similarly, for the inverse scattering $N (p_1,s) + N (p_2,r) \rightarrow a (p_3) + a (p_4)$ we have
\begin{align}
    \sigma_{NNaa} = \frac{4m_N^4}{\pi f_a^4 s^{3/2}}
    \Bigg[&-\frac{\sqrt{s-4m_N^2}(2m_a^4-8m_a^2 m_N^2 +m_N^2 s)}{m_a^4 - 4m_a^2 m_N^2 + m_N^2 s} \nonumber\\ 
    &+ \frac{2(s-2m_a^2)\rm{coth}^{-1}\left(\frac{s-2m_a^2}{\sqrt{s-4m_a^2}\sqrt{s-4m_N^2}}\right)}{\sqrt{s-4m_a^2}}\Bigg],
\end{align}
where $s \equiv (p_1+p_2)^2 = (p_3+p_4)^2.$ Likewise, for $aa \rightarrow \nu \nu$ scattering with t-channel HNL mediation,
\begin{align}
    \sigma_{aa\nu\nu} = &\frac{4m_N^4 \sqrt{s-4m_a^2}U_{eN}^4}{\pi f_a^4 s^{3/2}} \nonumber\\
    \times\Bigg[&-\frac{2(m_a^4 +m_N^4+m_\nu^4-2m_a^2m_N^2  -2m_a^2m_\nu^2-2m_N^2m_\nu^2)+m_N^2s}{m_a^4 +m_N^4+m_\nu^4-2m_a^2m_N^2  -2m_a^2m_\nu^2-2m_N^2m_\nu^2+m_N^2s} \nonumber \\
    &+\frac{2(2(m_a^2-m_N^2+m_\nu^2)-s)\rm{coth}^{-1}\left(\frac{2(m_a^2-m_N^2+m_\nu^2)-s}{\sqrt{s-4m_a^2}\sqrt{s-4m_\nu^2}}\right)}{\sqrt{s-4m_a^2}\sqrt{s-4m_\nu^2}}\Bigg].
\end{align}
In the limit, $m_\nu \rightarrow 0$, we will get simplified expression for $aa \rightarrow \nu\nu$ scattering as
\begin{align}
    \sigma_{aa\nu\nu} = \frac{4m_N^4\sqrt{s-4m_a^2}U_{eN}^4}{\pi f_a^4 s^{3/2}}
    \Bigg[&-\frac{2(m_a^4+m_N^4-2m_a^2m_N^2) +m_N^2 s}{m_a^4+m_N^4-2m_a^2m_N^2 +m_N^2 s} 
    \nonumber \\
    &+ \frac{2(2(m_a^2-m_N^2)-s)\rm{coth}^{-1}\left(\frac{2(m_a^2-m_N^2)-s}{\sqrt{s-4m_a^2}\sqrt{s}}\right)}{\sqrt{s-4m_a^2}\sqrt{s}}\Bigg].
\end{align}
The cross-section for $aa \rightarrow \nu N$ scattering (in the limit $m_\nu \rightarrow 0$) with t-channel HNL mediation,
\begin{align}
    \sigma_{aa\nu N} = \frac{2m_N^4\sqrt{s-4m_a^2}U_{eN}^2}{\pi f_a^4 s^{3/2}}
    \Bigg[&-\frac{2(2m_a^4-7m_a^2m_N^2+5m_N^4+m_N^2s)}{m_a^4-2m_a^2m_N^2+m_N^4+m_N^2s} 
    \nonumber\\
    &+\frac{4(2m_a^2-4m_N^2-s)\rm{coth}^{-1}\left(\frac{2(m_a^2-m_N^2)-s}{\sqrt{s-4m_a^2}\sqrt{s}}\right)}{\sqrt{s-4m_a^2}\sqrt{s}}\Bigg].
\end{align}
The cross-section for $aa \rightarrow NN$ scattering (in the limit $m_\nu \rightarrow 0$) with t-channel $\nu$ mediation,
\begin{align}
    \sigma_{aaNN} = \frac{4m_N^4 \sqrt{s-4m_a^2}U_{eN}^4}{\pi f_a^4 s^{3/2}}
    \Bigg[&-\frac{2m_a^4-4m_a^2m_N^2+4m_N^4-m_N^2s}{m_a^4-2m_a^2m_N^2+m_N^4} \nonumber \\
    &+\frac{2(2(m_a^2+m_N^2)-s)\rm{coth}^{-1}\left(\frac{2(m_a^2+m_N^2)-s}{\sqrt{s-4m_a^2}\sqrt{s-4m_N^2}}\right)}{\sqrt{s-4m_a^2}\sqrt{s-4m_N^2}}\Bigg].
\end{align}
The cross-section for $aa \rightarrow \nu\nu$ scattering (in the limit of $m_\nu \rightarrow 0$) with t-channel $\nu$ mediation,
\begin{align}
    \sigma_{aa\nu\nu} = \frac{4m_N^4\sqrt{s-4m_a^2}U_{eN}^8}{\pi f_a^4 s^{3/2}}
    \Bigg[-2+\frac{2(s-2m_a^2)\rm{coth}^{-1}\left(\frac{s-2m_a^2}{\sqrt{s-4m_a^2}\sqrt{s}}\right)}{\sqrt{s-4m_a^2}\sqrt{s}}\Bigg].
\end{align}
The cross-section for $aa \rightarrow \nu N$ scattering (in the limit $m_\nu \rightarrow 0$) with t-channel $\nu$ mediation,
\begin{align}
    \sigma_{aa\nu N} = \frac{4m_N^4 \sqrt{s-4m_a^2}U_{eN}^6}{\pi f_a^4 s^{3/2}}
    \Bigg[-2+\frac{m_N^2}{m_a^2} + \frac{2(s-2m_a^2)\rm{coth}^{-1}\left(\frac{s-2m_a^2}{\sqrt{s-4m_a^2}\sqrt{s}}\right)}{\sqrt{s-4m_a^2}\sqrt{s}}\Bigg].
\end{align}
The cross-section for $\nu \nu \rightarrow NN$ scattering (in the limit $m_\nu \rightarrow 0$) with t-channel axion mediation,
\begin{align}
    \sigma_{\nu\nu NN} = \frac{m_N^4 U_{eN}^4}{\pi f_a^4 s}
    \Bigg[&\frac{2m_a^4+2m_N^4+m_N^2s+2m_a^2(s-2m_N^2)}{m_a^4+m_N^4+m_a^2(s-2m_N^2)}
    \nonumber\\
    &+\frac{2(s+2m_a^2-2m_N^2)\rm{coth}^{-1}\left(\frac{-2m_a^2+2m_N^2-s}{\sqrt{s-4m_N^2}\sqrt{s}}\right)}{\sqrt{s-4m_N^2}\sqrt{s}}\Bigg].
\end{align}
The cross-section for $a\nu \rightarrow a N$ (in the limit $m_\nu \rightarrow 0$) with s and t-channel HNL mediation,
\begin{align}
    \sigma_{a\nu aN} = &\frac{m_N^4 U_{eN}^2}{2\pi f_a^4 (m_N^2-s)^2 s}
    \nonumber\\
    \times\Bigg[&\frac{1}{m_a^4-4m_a^2m_N^2+m_N^2s}
    \nonumber\\
    \Big\{&2m_a^8+m_N^2(8m_N^6-20m_N^4s+15m_N^2s^2+(5-2m_N^2)s^3 +2s^4)
    \nonumber \\
    &+2m_a^2(3m_N^6+6m_N^4(s-1)s-6m_N^2s^2(3+s))+2m_a^6(2m_N^2(s-3)
    \nonumber\\
    &-s(3+2s))+m_a^4(2m_N^4(7-8s)+s^2(7+2s)+m_N^2s(29+14s))\Big\}
    \nonumber\\
    &-4\sqrt{\frac{s-m_N^2}{s-4m_a^2}}\Big\{2m_a^4+m_N^4(3-2s)-s^2+2m_N^2s(s+3)-m_a^2(3s
    \nonumber\\
    &+m_N^2(3+4s))\Big\}\rm{coth}^{-1}\Bigg(\frac{s-2m_a^2}{\sqrt{s-4m_a^2}\sqrt{s-4m_N^2}}\Bigg)\Bigg].
\end{align}

\bibliographystyle{JHEP}
\bibliography{biblio}

\providecommand{\href}[2]{#2}\begingroup\raggedright\begin{thebibliography}{10}

\bibitem{Minkowski:1977sc}
P.~Minkowski, \emph{{$\mu \to e\gamma$ at a Rate of One Out of $10^{9}$ Muon
  Decays?}}, \href{https://doi.org/10.1016/0370-2693(77)90435-X}{\emph{Phys.
  Lett. B} {\bfseries 67} (1977) 421}.

\bibitem{Mohapatra:1979ia}
R.N.~Mohapatra and G.~Senjanovic, \emph{{Neutrino Mass and Spontaneous Parity
  Nonconservation}},
  \href{https://doi.org/10.1103/PhysRevLett.44.912}{\emph{Phys. Rev. Lett.}
  {\bfseries 44} (1980) 912}.

\bibitem{Gell-Mann:1979vob}
M.~Gell-Mann, P.~Ramond and R.~Slansky, \emph{{Complex Spinors and Unified
  Theories}}, {\emph{Conf. Proc. C} {\bfseries 790927} (1979) 315}
  [\href{https://arxiv.org/abs/1306.4669}{{\ttfamily 1306.4669}}].

\bibitem{Yanagida:1979as}
T.~Yanagida, \emph{{Horizontal gauge symmetry and masses of neutrinos}},
  {\emph{Conf. Proc. C} {\bfseries 7902131} (1979) 95}.

\bibitem{Schechter:1980gr}
J.~Schechter and J.W.F.~Valle, \emph{{Neutrino Masses in SU(2) $\times$ U(1)
  Theories}}, \href{https://doi.org/10.1103/PhysRevD.22.2227}{\emph{Phys. Rev.
  D} {\bfseries 22} (1980) 2227}.

\bibitem{Fukugita:1986hr}
M.~Fukugita and T.~Yanagida, \emph{{Baryogenesis Without Grand Unification}},
  \href{https://doi.org/10.1016/0370-2693(86)91126-3}{\emph{Phys. Lett. B}
  {\bfseries 174} (1986) 45}.

\bibitem{ParticleDataGroup:2024cfk}
{\scshape Particle Data Group} collaboration, \emph{{Review of particle
  physics}}, \href{https://doi.org/10.1103/PhysRevD.110.030001}{\emph{Phys.
  Rev. D} {\bfseries 110} (2024) 030001}.

\bibitem{Esteban:2020cvm}
I.~Esteban, M.C.~Gonzalez-Garcia, M.~Maltoni, T.~Schwetz and A.~Zhou,
  \emph{{The fate of hints: updated global analysis of three-flavor neutrino
  oscillations}}, \href{https://doi.org/10.1007/JHEP09(2020)178}{\emph{JHEP}
  {\bfseries 09} (2020) 178}
  [\href{https://arxiv.org/abs/2007.14792}{{\ttfamily 2007.14792}}].

\bibitem{Bolton:2019pcu}
P.D.~Bolton, F.F.~Deppisch and P.S.~Bhupal~Dev, \emph{{Neutrinoless double beta
  decay versus other probes of heavy sterile neutrinos}},
  \href{https://doi.org/10.1007/JHEP03(2020)170}{\emph{JHEP} {\bfseries 03}
  (2020) 170} [\href{https://arxiv.org/abs/1912.03058}{{\ttfamily
  1912.03058}}].

\bibitem{Abdullahi:2022jlv}
A.M.~Abdullahi et~al., \emph{{The present and future status of heavy neutral
  leptons}}, \href{https://doi.org/10.1088/1361-6471/ac98f9}{\emph{J. Phys. G}
  {\bfseries 50} (2023) 020501}
  [\href{https://arxiv.org/abs/2203.08039}{{\ttfamily 2203.08039}}].

\bibitem{Chrzaszcz:2019inj}
M.~Chrzaszcz, M.~Drewes, T.E.~Gonzalo, J.~Harz, S.~Krishnamurthy and
  C.~Weniger, \emph{{A frequentist analysis of three right-handed neutrinos
  with GAMBIT}},
  \href{https://doi.org/10.1140/epjc/s10052-020-8073-9}{\emph{Eur. Phys. J. C}
  {\bfseries 80} (2020) 569}
  [\href{https://arxiv.org/abs/1908.02302}{{\ttfamily 1908.02302}}].

\bibitem{Boyarsky:2021yoh}
A.~Boyarsky, M.~Ovchynnikov, N.~Sabti and V.~Syvolap, \emph{{When feebly
  interacting massive particles decay into neutrinos: The Neff story}},
  \href{https://doi.org/10.1103/PhysRevD.104.035006}{\emph{Phys. Rev. D}
  {\bfseries 104} (2021) 035006}
  [\href{https://arxiv.org/abs/2103.09831}{{\ttfamily 2103.09831}}].

\bibitem{GAMBITCosmologyWorkgroup:2020htv}
{\scshape GAMBIT Cosmology Workgroup} collaboration, \emph{{CosmoBit: A GAMBIT
  module for computing cosmological observables and likelihoods}},
  \href{https://doi.org/10.1088/1475-7516/2021/02/022}{\emph{JCAP} {\bfseries
  02} (2021) 022} [\href{https://arxiv.org/abs/2009.03286}{{\ttfamily
  2009.03286}}].

\bibitem{Atre:2009rg}
A.~Atre, T.~Han, S.~Pascoli and B.~Zhang, \emph{{The Search for Heavy Majorana
  Neutrinos}}, \href{https://doi.org/10.1088/1126-6708/2009/05/030}{\emph{JHEP}
  {\bfseries 05} (2009) 030} [\href{https://arxiv.org/abs/0901.3589}{{\ttfamily
  0901.3589}}].

\bibitem{Coloma:2020lgy}
P.~Coloma, E.~Fern\'andez-Mart\'\i{}nez, M.~Gonz\'alez-L\'opez,
  J.~Hern\'andez-Garc\'\i{}a and Z.~Pavlovic, \emph{{GeV-scale neutrinos:
  interactions with mesons and DUNE sensitivity}},
  \href{https://doi.org/10.1140/epjc/s10052-021-08861-y}{\emph{Eur. Phys. J. C}
  {\bfseries 81} (2021) 78} [\href{https://arxiv.org/abs/2007.03701}{{\ttfamily
  2007.03701}}].

\bibitem{Hsyu:2020uqb}
T.~Hsyu, R.J.~Cooke, J.X.~Prochaska and M.~Bolte, \emph{{The PHLEK Survey: A
  New Determination of the Primordial Helium Abundance}},
  \href{https://doi.org/10.3847/1538-4357/ab91af}{\emph{Astrophys. J.}
  {\bfseries 896} (2020) 77}
  [\href{https://arxiv.org/abs/2005.12290}{{\ttfamily 2005.12290}}].

\bibitem{Davidson:1978pm}
A.~Davidson, \emph{{$B-L$ as the fourth color within an $\mathrm{SU}(2)_L
  \times \mathrm{U}(1)_R \times \mathrm{U}(1)$ model}},
  \href{https://doi.org/10.1103/PhysRevD.20.776}{\emph{Phys. Rev. D} {\bfseries
  20} (1979) 776}.

\bibitem{Marshak:1979fm}
R.E.~Marshak and R.N.~Mohapatra, \emph{{Quark - Lepton Symmetry and B-L as the
  U(1) Generator of the Electroweak Symmetry Group}},
  \href{https://doi.org/10.1016/0370-2693(80)90436-0}{\emph{Phys. Lett. B}
  {\bfseries 91} (1980) 222}.

\bibitem{Mohapatra:1980qe}
R.N.~Mohapatra and R.E.~Marshak, \emph{{Local B-L Symmetry of Electroweak
  Interactions, Majorana Neutrinos and Neutron Oscillations}},
  \href{https://doi.org/10.1103/PhysRevLett.44.1316}{\emph{Phys. Rev. Lett.}
  {\bfseries 44} (1980) 1316}.

\bibitem{Davidson:1987mh}
A.~Davidson and K.C.~Wali, \emph{{Universal Seesaw Mechanism?}},
  \href{https://doi.org/10.1103/PhysRevLett.59.393}{\emph{Phys. Rev. Lett.}
  {\bfseries 59} (1987) 393}.

\bibitem{Buchmuller:1991ce}
W.~Buchmuller, C.~Greub and P.~Minkowski, \emph{{Neutrino masses, neutral
  vector bosons and the scale of B-L breaking}},
  \href{https://doi.org/10.1016/0370-2693(91)90952-M}{\emph{Phys. Lett. B}
  {\bfseries 267} (1991) 395}.

\bibitem{Pati:1973rp}
J.C.~Pati and A.~Salam, \emph{{Is Baryon Number Conserved?}},
  \href{https://doi.org/10.1103/PhysRevLett.31.661}{\emph{Phys. Rev. Lett.}
  {\bfseries 31} (1973) 661}.

\bibitem{Pati:1973uk}
J.C.~Pati and A.~Salam, \emph{{Unified Lepton-Hadron Symmetry and a Gauge
  Theory of the Basic Interactions}},
  \href{https://doi.org/10.1103/PhysRevD.8.1240}{\emph{Phys. Rev. D} {\bfseries
  8} (1973) 1240}.

\bibitem{Pati:1974yy}
J.C.~Pati and A.~Salam, \emph{{Lepton Number as the Fourth Color}},
  \href{https://doi.org/10.1103/PhysRevD.10.275}{\emph{Phys. Rev. D} {\bfseries
  10} (1974) 275}.

\bibitem{Mohapatra:1974gc}
R.N.~Mohapatra and J.C.~Pati, \emph{{A Natural Left-Right Symmetry}},
  \href{https://doi.org/10.1103/PhysRevD.11.2558}{\emph{Phys. Rev. D}
  {\bfseries 11} (1975) 2558}.

\bibitem{Senjanovic:1975rk}
G.~Senjanovic and R.N.~Mohapatra, \emph{{Exact Left-Right Symmetry and
  Spontaneous Violation of Parity}},
  \href{https://doi.org/10.1103/PhysRevD.12.1502}{\emph{Phys. Rev. D}
  {\bfseries 12} (1975) 1502}.

\bibitem{Marsh:2015xka}
D.J.E.~Marsh, \emph{{Axion Cosmology}},
  \href{https://doi.org/10.1016/j.physrep.2016.06.005}{\emph{Phys. Rept.}
  {\bfseries 643} (2016) 1} [\href{https://arxiv.org/abs/1510.07633}{{\ttfamily
  1510.07633}}].

\bibitem{Chadha-Day:2021szb}
F.~Chadha-Day, J.~Ellis and D.J.E.~Marsh, \emph{{Axion dark matter: What is it
  and why now?}}, \href{https://doi.org/10.1126/sciadv.abj3618}{\emph{Sci.
  Adv.} {\bfseries 8} (2022) abj3618}
  [\href{https://arxiv.org/abs/2105.01406}{{\ttfamily 2105.01406}}].

\bibitem{Marsh:2023tep}
D.J.E.~Marsh, \emph{{Axions for amateurs}},
  \href{https://doi.org/10.1080/00107514.2023.2256085}{\emph{Contemp. Phys.}
  {\bfseries 64} (2023) 1} [\href{https://arxiv.org/abs/2308.16003}{{\ttfamily
  2308.16003}}].

\bibitem{deGiorgi:2022oks}
A.~de~Giorgi, L.~Merlo and J.-L.~Tastet, \emph{{Probing HNL-ALP couplings at
  colliders}}, \href{https://doi.org/10.1002/prop.202300027}{\emph{Fortsch.
  Phys.} {\bfseries 71} (2023) 2300027}
  [\href{https://arxiv.org/abs/2212.11290}{{\ttfamily 2212.11290}}].

\bibitem{Marcos:2024yfm}
M.B.~Marcos, A.~de~Giorgi, L.~Merlo and J.-L.~Tastet, \emph{{ALPs and HNLs at
  LHC and Muon Colliders: Uncovering New Couplings and Signals}},
  \href{https://arxiv.org/abs/2407.14970}{{\ttfamily 2407.14970}}.

\bibitem{Wang:2024mrc}
Z.S.~Wang, Y.~Zhang and W.~Liu, \emph{{Searching for heavy neutral leptons
  coupled to axion-like particles at the LHC far detectors and SHiP}},
  \href{https://arxiv.org/abs/2409.18424}{{\ttfamily 2409.18424}}.

\bibitem{Wang:2024prt}
Z.S.~Wang, Y.~Zhang and W.~Liu, \emph{{Long-lived sterile neutrinos from an
  axionlike particle at Belle II}},
  \href{https://arxiv.org/abs/2410.00491}{{\ttfamily 2410.00491}}.

\bibitem{Arguelles:2021dqn}
C.A.~Arg\"uelles, N.~Foppiani and M.~Hostert, \emph{{Heavy neutral leptons
  below the kaon mass at hodoscopic neutrino detectors}},
  \href{https://doi.org/10.1103/PhysRevD.105.095006}{\emph{Phys. Rev. D}
  {\bfseries 105} (2022) 095006}
  [\href{https://arxiv.org/abs/2109.03831}{{\ttfamily 2109.03831}}].

\bibitem{Gola:2021abm}
S.~Gola, S.~Mandal and N.~Sinha, \emph{{ALP-portal majorana dark matter}},
  \href{https://doi.org/10.1142/S0217751X22501317}{\emph{Int. J. Mod. Phys. A}
  {\bfseries 37} (2022) 2250131}
  [\href{https://arxiv.org/abs/2106.00547}{{\ttfamily 2106.00547}}].

\bibitem{Gelmini:1982zz}
G.B.~Gelmini, S.~Nussinov and T.~Yanagida, \emph{{Does Nature Like
  Nambu-Goldstone Bosons?}},
  \href{https://doi.org/10.1016/0550-3213(83)90426-1}{\emph{Nucl. Phys. B}
  {\bfseries 219} (1983) 31}.

\bibitem{Gondolo:1990dk}
P.~Gondolo and G.~Gelmini, \emph{{Cosmic abundances of stable particles:
  Improved analysis}},
  \href{https://doi.org/10.1016/0550-3213(91)90438-4}{\emph{Nucl. Phys. B}
  {\bfseries 360} (1991) 145}.

\bibitem{Escudero:2018mvt}
M.~Escudero, \emph{{Neutrino decoupling beyond the Standard Model: CMB
  constraints on the Dark Matter mass with a fast and precise $N_{\rm eff}$
  evaluation}},
  \href{https://doi.org/10.1088/1475-7516/2019/02/007}{\emph{JCAP} {\bfseries
  02} (2019) 007} [\href{https://arxiv.org/abs/1812.05605}{{\ttfamily
  1812.05605}}].

\bibitem{Hufnagel:2018bjp}
M.~Hufnagel, K.~Schmidt-Hoberg and S.~Wild, \emph{{BBN constraints on MeV-scale
  dark sectors. Part II. Electromagnetic decays}},
  \href{https://doi.org/10.1088/1475-7516/2018/11/032}{\emph{JCAP} {\bfseries
  11} (2018) 032} [\href{https://arxiv.org/abs/1808.09324}{{\ttfamily
  1808.09324}}].

\bibitem{EscuderoAbenza:2020cmq}
M.~Escudero~Abenza, \emph{{Precision early universe thermodynamics made simple:
  $N_{\rm eff}$ and neutrino decoupling in the Standard Model and beyond}},
  \href{https://doi.org/10.1088/1475-7516/2020/05/048}{\emph{JCAP} {\bfseries
  05} (2020) 048} [\href{https://arxiv.org/abs/2001.04466}{{\ttfamily
  2001.04466}}].

\bibitem{Cadamuro:2011fd}
D.~Cadamuro and J.~Redondo, \emph{{Cosmological bounds on pseudo
  Nambu-Goldstone bosons}},
  \href{https://doi.org/10.1088/1475-7516/2012/02/032}{\emph{JCAP} {\bfseries
  02} (2012) 032} [\href{https://arxiv.org/abs/1110.2895}{{\ttfamily
  1110.2895}}].

\bibitem{Drewes:2015iva}
M.~Drewes and B.~Garbrecht, \emph{{Combining experimental and cosmological
  constraints on heavy neutrinos}},
  \href{https://doi.org/10.1016/j.nuclphysb.2017.05.001}{\emph{Nucl. Phys. B}
  {\bfseries 921} (2017) 250}
  [\href{https://arxiv.org/abs/1502.00477}{{\ttfamily 1502.00477}}].

\bibitem{Depta:2020zbh}
P.F.~Depta, M.~Hufnagel and K.~Schmidt-Hoberg, \emph{{Updated BBN constraints
  on electromagnetic decays of MeV-scale particles}},
  \href{https://doi.org/10.1088/1475-7516/2021/04/011}{\emph{JCAP} {\bfseries
  04} (2021) 011} [\href{https://arxiv.org/abs/2011.06519}{{\ttfamily
  2011.06519}}].

\bibitem{Balazs:2022tjl}
C.~Bal\'azs et~al., \emph{{Cosmological constraints on decaying axion-like
  particles: a global analysis}},
  \href{https://doi.org/10.1088/1475-7516/2022/12/027}{\emph{JCAP} {\bfseries
  12} (2022) 027} [\href{https://arxiv.org/abs/2205.13549}{{\ttfamily
  2205.13549}}].

\bibitem{Iocco:2008va}
F.~Iocco, G.~Mangano, G.~Miele, O.~Pisanti and P.D.~Serpico, \emph{{Primordial
  Nucleosynthesis: from precision cosmology to fundamental physics}},
  \href{https://doi.org/10.1016/j.physrep.2009.02.002}{\emph{Phys. Rept.}
  {\bfseries 472} (2009) 1} [\href{https://arxiv.org/abs/0809.0631}{{\ttfamily
  0809.0631}}].

\bibitem{Coc:2017pxv}
A.~Coc and E.~Vangioni, \emph{{Primordial nucleosynthesis}},
  \href{https://doi.org/10.1142/S0218301317410026}{\emph{Int. J. Mod. Phys. E}
  {\bfseries 26} (2017) 1741002}
  [\href{https://arxiv.org/abs/1707.01004}{{\ttfamily 1707.01004}}].

\bibitem{Boyarsky:2020dzc}
A.~Boyarsky, M.~Ovchynnikov, O.~Ruchayskiy and V.~Syvolap, \emph{{Improved big
  bang nucleosynthesis constraints on heavy neutral leptons}},
  \href{https://doi.org/10.1103/PhysRevD.104.023517}{\emph{Phys. Rev. D}
  {\bfseries 104} (2021) 023517}
  [\href{https://arxiv.org/abs/2008.00749}{{\ttfamily 2008.00749}}].

\bibitem{Kohri:2001jx}
K.~Kohri, \emph{{Primordial nucleosynthesis and hadronic decay of a massive
  particle with a relatively short lifetime}},
  \href{https://doi.org/10.1103/PhysRevD.64.043515}{\emph{Phys. Rev. D}
  {\bfseries 64} (2001) 043515}
  [\href{https://arxiv.org/abs/astro-ph/0103411}{{\ttfamily
  astro-ph/0103411}}].

\bibitem{Kawasaki:2004qu}
M.~Kawasaki, K.~Kohri and T.~Moroi, \emph{{Big-Bang nucleosynthesis and
  hadronic decay of long-lived massive particles}},
  \href{https://doi.org/10.1103/PhysRevD.71.083502}{\emph{Phys. Rev. D}
  {\bfseries 71} (2005) 083502}
  [\href{https://arxiv.org/abs/astro-ph/0408426}{{\ttfamily
  astro-ph/0408426}}].

\bibitem{Kanzaki:2007pd}
T.~Kanzaki, M.~Kawasaki, K.~Kohri and T.~Moroi, \emph{{Cosmological Constraints
  on Neutrino Injection}},
  \href{https://doi.org/10.1103/PhysRevD.76.105017}{\emph{Phys. Rev. D}
  {\bfseries 76} (2007) 105017}
  [\href{https://arxiv.org/abs/0705.1200}{{\ttfamily 0705.1200}}].

\bibitem{Kawasaki:2000en}
M.~Kawasaki, K.~Kohri and N.~Sugiyama, \emph{{MeV scale reheating temperature
  and thermalization of neutrino background}},
  \href{https://doi.org/10.1103/PhysRevD.62.023506}{\emph{Phys. Rev. D}
  {\bfseries 62} (2000) 023506}
  [\href{https://arxiv.org/abs/astro-ph/0002127}{{\ttfamily
  astro-ph/0002127}}].

\bibitem{Hasegawa:2019jsa}
T.~Hasegawa, N.~Hiroshima, K.~Kohri, R.S.L.~Hansen, T.~Tram and S.~Hannestad,
  \emph{{MeV-scale reheating temperature and thermalization of oscillating
  neutrinos by radiative and hadronic decays of massive particles}},
  \href{https://doi.org/10.1088/1475-7516/2019/12/012}{\emph{JCAP} {\bfseries
  12} (2019) 012} [\href{https://arxiv.org/abs/1908.10189}{{\ttfamily
  1908.10189}}].

\bibitem{Dolgov:2000jw}
A.D.~Dolgov, S.H.~Hansen, G.~Raffelt and D.V.~Semikoz, \emph{{Heavy sterile
  neutrinos: Bounds from big bang nucleosynthesis and SN1987A}},
  \href{https://doi.org/10.1016/S0550-3213(00)00566-6}{\emph{Nucl. Phys. B}
  {\bfseries 590} (2000) 562}
  [\href{https://arxiv.org/abs/hep-ph/0008138}{{\ttfamily hep-ph/0008138}}].

\bibitem{Ruchayskiy:2012si}
O.~Ruchayskiy and A.~Ivashko, \emph{{Restrictions on the lifetime of sterile
  neutrinos from primordial nucleosynthesis}},
  \href{https://doi.org/10.1088/1475-7516/2012/10/014}{\emph{JCAP} {\bfseries
  10} (2012) 014} [\href{https://arxiv.org/abs/1202.2841}{{\ttfamily
  1202.2841}}].

\bibitem{Sabti:2020yrt}
N.~Sabti, A.~Magalich and A.~Filimonova, \emph{{An Extended Analysis of Heavy
  Neutral Leptons during Big Bang Nucleosynthesis}},
  \href{https://doi.org/10.1088/1475-7516/2020/11/056}{\emph{JCAP} {\bfseries
  11} (2020) 056} [\href{https://arxiv.org/abs/2006.07387}{{\ttfamily
  2006.07387}}].

\bibitem{Kawasaki:1994bs}
M.~Kawasaki and T.~Moroi, \emph{{Gravitino decay into a neutrino and a
  sneutrino in the inflationary universe}},
  \href{https://doi.org/10.1016/0370-2693(94)01644-R}{\emph{Phys. Lett. B}
  {\bfseries 346} (1995) 27}
  [\href{https://arxiv.org/abs/hep-ph/9408321}{{\ttfamily hep-ph/9408321}}].

\bibitem{Kawasaki:2004yh}
M.~Kawasaki, K.~Kohri and T.~Moroi, \emph{{Hadronic decay of late - decaying
  particles and Big-Bang Nucleosynthesis}},
  \href{https://doi.org/10.1016/j.physletb.2005.08.045}{\emph{Phys. Lett. B}
  {\bfseries 625} (2005) 7}
  [\href{https://arxiv.org/abs/astro-ph/0402490}{{\ttfamily
  astro-ph/0402490}}].

\bibitem{Kanzaki:2006hm}
T.~Kanzaki, M.~Kawasaki, K.~Kohri and T.~Moroi, \emph{{Cosmological constraints
  on gravitino LSP scenario with sneutrino NLSP}},
  \href{https://doi.org/10.1103/PhysRevD.75.025011}{\emph{Phys. Rev. D}
  {\bfseries 75} (2007) 025011}
  [\href{https://arxiv.org/abs/hep-ph/0609246}{{\ttfamily hep-ph/0609246}}].

\bibitem{Jedamzik:2006xz}
K.~Jedamzik, \emph{{Big bang nucleosynthesis constraints on hadronically and
  electromagnetically decaying relic neutral particles}},
  \href{https://doi.org/10.1103/PhysRevD.74.103509}{\emph{Phys. Rev. D}
  {\bfseries 74} (2006) 103509}
  [\href{https://arxiv.org/abs/hep-ph/0604251}{{\ttfamily hep-ph/0604251}}].

\bibitem{Ishiwata:2009gs}
K.~Ishiwata, M.~Kawasaki, K.~Kohri and T.~Moroi, \emph{{Right-handed sneutrino
  dark matter and big-bang nucleosynthesis}},
  \href{https://doi.org/10.1016/j.physletb.2010.04.054}{\emph{Phys. Lett. B}
  {\bfseries 689} (2010) 163}
  [\href{https://arxiv.org/abs/0912.0781}{{\ttfamily 0912.0781}}].

\bibitem{Planck:2018vyg}
{\scshape Planck} collaboration, \emph{{Planck 2018 results. VI. Cosmological
  parameters}},
  \href{https://doi.org/10.1051/0004-6361/201833910}{\emph{Astron. Astrophys.}
  {\bfseries 641} (2020) A6}
  [\href{https://arxiv.org/abs/1807.06209}{{\ttfamily 1807.06209}}].

\bibitem{Fields:2019pfx}
B.D.~Fields, K.A.~Olive, T.-H.~Yeh and C.~Young, \emph{{Big-Bang
  Nucleosynthesis after Planck}},
  \href{https://doi.org/10.1088/1475-7516/2020/03/010}{\emph{JCAP} {\bfseries
  03} (2020) 010} [\href{https://arxiv.org/abs/1912.01132}{{\ttfamily
  1912.01132}}].

\bibitem{Kamiokande-II:1987idp}
{\scshape Kamiokande-II} collaboration, \emph{{Observation of a Neutrino Burst
  from the Supernova SN 1987a}},
  \href{https://doi.org/10.1103/PhysRevLett.58.1490}{\emph{Phys. Rev. Lett.}
  {\bfseries 58} (1987) 1490}.

\bibitem{Totsuka:1988iyh}
Y.~Totsuka et~al., \emph{{Observation of a neutrino burst from the supernova
  SN1987a}}, \href{https://doi.org/10.1016/0375-9474(88)90844-5}{\emph{Nucl.
  Phys. A} {\bfseries 478} (1988) 189}.

\bibitem{Dolgov:2000pj}
A.D.~Dolgov, S.H.~Hansen, G.~Raffelt and D.V.~Semikoz, \emph{{Cosmological and
  astrophysical bounds on a heavy sterile neutrino and the KARMEN anomaly}},
  \href{https://doi.org/10.1016/S0550-3213(00)00203-0}{\emph{Nucl. Phys. B}
  {\bfseries 580} (2000) 331}
  [\href{https://arxiv.org/abs/hep-ph/0002223}{{\ttfamily hep-ph/0002223}}].

\bibitem{Fuller:2008erj}
G.M.~Fuller, A.~Kusenko and K.~Petraki, \emph{{Heavy sterile neutrinos and
  supernova explosions}},
  \href{https://doi.org/10.1016/j.physletb.2008.11.016}{\emph{Phys. Lett. B}
  {\bfseries 670} (2009) 281}
  [\href{https://arxiv.org/abs/0806.4273}{{\ttfamily 0806.4273}}].

\bibitem{Mastrototaro:2019vug}
L.~Mastrototaro, A.~Mirizzi, P.D.~Serpico and A.~Esmaili, \emph{{Heavy sterile
  neutrino emission in core-collapse supernovae: Constraints and signatures}},
  \href{https://doi.org/10.1088/1475-7516/2020/01/010}{\emph{JCAP} {\bfseries
  01} (2020) 010} [\href{https://arxiv.org/abs/1910.10249}{{\ttfamily
  1910.10249}}].

\bibitem{Fiorillo:2022cdq}
D.F.G.~Fiorillo, G.G.~Raffelt and E.~Vitagliano, \emph{{Strong Supernova 1987A
  Constraints on Bosons Decaying to Neutrinos}},
  \href{https://doi.org/10.1103/PhysRevLett.131.021001}{\emph{Phys. Rev. Lett.}
  {\bfseries 131} (2023) 021001}
  [\href{https://arxiv.org/abs/2209.11773}{{\ttfamily 2209.11773}}].

\bibitem{Syvolap:2023trc}
V.~Syvolap, \emph{{Testing heavy neutral leptons produced in the supernovae
  explosions with future neutrino detectors}},
  \href{https://arxiv.org/abs/2301.07052}{{\ttfamily 2301.07052}}.

\bibitem{Akita:2023iwq}
K.~Akita, S.H.~Im, M.~Masud and S.~Yun, \emph{{Limits on heavy neutral leptons,
  $Z^\prime$ bosons and majorons from high-energy supernova neutrinos}},
  \href{https://doi.org/10.1007/JHEP07(2024)057}{\emph{JHEP} {\bfseries 07}
  (2024) 057} [\href{https://arxiv.org/abs/2312.13627}{{\ttfamily
  2312.13627}}].

\bibitem{Kainulainen:1990bn}
K.~Kainulainen, J.~Maalampi and J.T.~Peltoniemi, \emph{{Inert neutrinos in
  supernovae}}, \href{https://doi.org/10.1016/0550-3213(91)90354-Z}{\emph{Nucl.
  Phys. B} {\bfseries 358} (1991) 435}.

\bibitem{Raffelt:1992bs}
G.~Raffelt and G.~Sigl, \emph{{Neutrino flavor conversion in a supernova
  core}}, \href{https://doi.org/10.1016/0927-6505(93)90020-E}{\emph{Astropart.
  Phys.} {\bfseries 1} (1993) 165}
  [\href{https://arxiv.org/abs/astro-ph/9209005}{{\ttfamily
  astro-ph/9209005}}].

\bibitem{Telalovic:2024cot}
B.~Telalovic, D.F.G.~Fiorillo, P.~Mart\'\i{}nez-Mirav\'e, E.~Vitagliano and
  M.~Bustamante, \emph{{The next galactic supernova can uncover mass and
  couplings of particles decaying to neutrinos}},
  \href{https://arxiv.org/abs/2406.15506}{{\ttfamily 2406.15506}}.

\bibitem{Jaeckel:2017tud}
J.~Jaeckel, P.C.~Malta and J.~Redondo, \emph{{Decay photons from the axionlike
  particles burst of type II supernovae}},
  \href{https://doi.org/10.1103/PhysRevD.98.055032}{\emph{Phys. Rev. D}
  {\bfseries 98} (2018) 055032}
  [\href{https://arxiv.org/abs/1702.02964}{{\ttfamily 1702.02964}}].

\bibitem{Hoof:2022xbe}
S.~Hoof and L.~Schulz, \emph{{Updated constraints on axion-like particles from
  temporal information in supernova SN1987A gamma-ray data}},
  \href{https://doi.org/10.1088/1475-7516/2023/03/054}{\emph{JCAP} {\bfseries
  03} (2023) 054} [\href{https://arxiv.org/abs/2212.09764}{{\ttfamily
  2212.09764}}].

\bibitem{Diamond:2023scc}
M.~Diamond, D.F.G.~Fiorillo, G.~Marques-Tavares and E.~Vitagliano,
  \emph{{Axion-sourced fireballs from supernovae}},
  \href{https://doi.org/10.1103/PhysRevD.107.103029}{\emph{Phys. Rev. D}
  {\bfseries 107} (2023) 103029}
  [\href{https://arxiv.org/abs/2303.11395}{{\ttfamily 2303.11395}}].

\bibitem{Diamond:2023cto}
M.~Diamond, D.F.G.~Fiorillo, G.~Marques-Tavares, I.~Tamborra and E.~Vitagliano,
  \emph{{Multimessenger Constraints on Radiatively Decaying Axions from
  GW170817}}, \href{https://doi.org/10.1103/PhysRevLett.132.101004}{\emph{Phys.
  Rev. Lett.} {\bfseries 132} (2024) 101004}
  [\href{https://arxiv.org/abs/2305.10327}{{\ttfamily 2305.10327}}].

\bibitem{Caputo:2022mah}
A.~Caputo, H.-T.~Janka, G.~Raffelt and E.~Vitagliano, \emph{{Low-Energy
  Supernovae Severely Constrain Radiative Particle Decays}},
  \href{https://doi.org/10.1103/PhysRevLett.128.221103}{\emph{Phys. Rev. Lett.}
  {\bfseries 128} (2022) 221103}
  [\href{https://arxiv.org/abs/2201.09890}{{\ttfamily 2201.09890}}].

\bibitem{Caputo:2021rux}
A.~Caputo, G.~Raffelt and E.~Vitagliano, \emph{{Muonic boson limits: Supernova
  redux}}, \href{https://doi.org/10.1103/PhysRevD.105.035022}{\emph{Phys. Rev.
  D} {\bfseries 105} (2022) 035022}
  [\href{https://arxiv.org/abs/2109.03244}{{\ttfamily 2109.03244}}].

\bibitem{Giannotti:2015kwo}
M.~Giannotti, I.~Irastorza, J.~Redondo and A.~Ringwald, \emph{{Cool WISPs for
  stellar cooling excesses}},
  \href{https://doi.org/10.1088/1475-7516/2016/05/057}{\emph{JCAP} {\bfseries
  05} (2016) 057} [\href{https://arxiv.org/abs/1512.08108}{{\ttfamily
  1512.08108}}].

\bibitem{DiLuzio:2020wdo}
L.~Di~Luzio, M.~Giannotti, E.~Nardi and L.~Visinelli, \emph{{The landscape of
  QCD axion models}},
  \href{https://doi.org/10.1016/j.physrep.2020.06.002}{\emph{Phys. Rept.}
  {\bfseries 870} (2020) 1} [\href{https://arxiv.org/abs/2003.01100}{{\ttfamily
  2003.01100}}].

\bibitem{PIONEER:2022yag}
{\scshape PIONEER} collaboration, \emph{{PIONEER: Studies of Rare Pion
  Decays}},  \href{https://arxiv.org/abs/2203.01981}{{\ttfamily 2203.01981}}.

\bibitem{Dias:2022fwf}
K.~Dias, \emph{{Search for Heavy Neutral Lepton Production in
  NA62${}^{\mathbf{\#}}$}},
  \href{https://doi.org/10.3103/S0027134922020497}{\emph{Moscow Univ. Phys.
  Bull.} {\bfseries 77} (2022) 220}.

\bibitem{Drewes:2018gkc}
M.~Drewes, J.~Hajer, J.~Klaric and G.~Lanfranchi, \emph{{NA62 sensitivity to
  heavy neutral leptons in the low scale seesaw model}},
  \href{https://doi.org/10.1007/JHEP07(2018)105}{\emph{JHEP} {\bfseries 07}
  (2018) 105} [\href{https://arxiv.org/abs/1801.04207}{{\ttfamily
  1801.04207}}].

\bibitem{Ballett:2019bgd}
P.~Ballett, T.~Boschi and S.~Pascoli, \emph{{Heavy Neutral Leptons from
  low-scale seesaws at the DUNE Near Detector}},
  \href{https://doi.org/10.1007/JHEP03(2020)111}{\emph{JHEP} {\bfseries 03}
  (2020) 111} [\href{https://arxiv.org/abs/1905.00284}{{\ttfamily
  1905.00284}}].

\bibitem{SHiP:2018xqw}
{\scshape SHiP} collaboration, \emph{{Sensitivity of the SHiP experiment to
  Heavy Neutral Leptons}},
  \href{https://doi.org/10.1007/JHEP04(2019)077}{\emph{JHEP} {\bfseries 04}
  (2019) 077} [\href{https://arxiv.org/abs/1811.00930}{{\ttfamily
  1811.00930}}].

\bibitem{Blondel:2022qqo}
A.~Blondel et~al., \emph{{Searches for long-lived particles at the future
  FCC-ee}}, \href{https://doi.org/10.3389/fphy.2022.967881}{\emph{Front. in
  Phys.} {\bfseries 10} (2022) 967881}
  [\href{https://arxiv.org/abs/2203.05502}{{\ttfamily 2203.05502}}].

\bibitem{Bolton:2022tds}
P.D.~Bolton, F.F.~Deppisch, M.~Rai and Z.~Zhang, \emph{{Probing the Nature of
  Heavy Neutral Leptons in Direct Searches and Neutrinoless Double Beta
  Decay}},  \href{https://arxiv.org/abs/2212.14690}{{\ttfamily 2212.14690}}.

\bibitem{AbdusSalam:2020rdj}
S.S.~AbdusSalam et~al., \emph{{Simple and statistically sound recommendations
  for analysing physical theories}},
  \href{https://doi.org/10.1088/1361-6633/ac60ac}{\emph{Rept. Prog. Phys.}
  {\bfseries 85} (2022) 052201}
  [\href{https://arxiv.org/abs/2012.09874}{{\ttfamily 2012.09874}}].

\bibitem{Carenza:2023old}
P.~Carenza, G.~Lucente, L.~Mastrototaro, A.~Mirizzi and P.D.~Serpico,
  \emph{{Comprehensive constraints on heavy sterile neutrinos from
  core-collapse supernovae}},
  \href{https://doi.org/10.1103/PhysRevD.109.063010}{\emph{Phys. Rev. D}
  {\bfseries 109} (2024) 063010}
  [\href{https://arxiv.org/abs/2311.00033}{{\ttfamily 2311.00033}}].

\end{thebibliography}\endgroup
\end{document}